\documentclass{aa}

\usepackage{natbib}
\usepackage{color}
\usepackage{tabularx}
\usepackage{amsmath, amssymb}
\usepackage{bm}
\usepackage{pifont}
\usepackage{float}
\usepackage{graphicx}
\usepackage[usenames,dvipsnames]{xcolor}
 \usepackage{relsize}
\usepackage{subfigure}  %inclusion de varias imagenes
\usepackage[singlelinecheck=false % <-- important
]{caption}

\usepackage{soul} %strikeoutcommand \st{text}
\usepackage{epsfig}
\usepackage{enumerate}
\usepackage[shortlabels]{enumitem}
\usepackage[version=3]{mhchem}
\usepackage{caption}
\usepackage{varwidth}
\usepackage{dcolumn}
\newsavebox\CBox
\usepackage{color}
\usepackage{lineno}
\usepackage{caption}
\usepackage{multirow}
\usepackage{epsf,epsfig}
\usepackage{epstopdf} % with I can solve the problem about the pictures.
\usepackage{amsmath,amssymb,pifont,rotate}
\usepackage{float}
\usepackage{subfloat}
\usepackage{rotating} 
     \def\aap{A\&A}
     \def\apjl{ApJL}
     \def\apjs{ApJS}

\begin{document}

\title{Environmental effects on the ionization\\ of brown dwarf
  atmospheres}
\titlerunning{Environmental effects on  brown dwarf atmospheres}

\author{M.\ I.\ Rodr\'iguez-Barrera\inst{1,2}, Ch.\ Helling\inst{1,2,3}\thanks{\email{ch80@st-andrews.ac.uk}}, \and K.\ Wood\inst{2}
           }
 \institute{
 $^1$ Centre for Exoplanet Science, University of St Andrews, UK \\
   $^2$ School of Physics \& Astronomy, University of St.\ Andrews, St.\ Andrews KY16 9SS, UK\\
$^3$ Kapteyn Astronomical Institute, Postbus 800, 9700 AV Groningen, The Netherlands
 }

\abstract
  % context heading (optional) 
{Brown dwarfs emit bursts of H$\alpha$, white light flares, and show radio flares and quiescent  radio emission. They are suggested to form Aurorae, similar to planets in the solar system but much more energetic. All these processes require a source gas with an appropriate degree of ionisation which, so far, is mostly postulated to be sufficient.}
%These observations of  ultra-cool objects suggest that the gas in their uppermost atmospheres is heated, ionised and magnetised to levels that radio and X-ray emission are possible. }
% aims heading (mandatory)
   {We aim to demonstrate that the galactic environment influences atmospheric ionisation, and that it  hence amplifies or enables the magnetic coupling of the atmospheres of ultra-cool objects, like brown dwarfs and free-floating planets.}
% methods heading (mandatory) 
   {We build on our previous work on thermal ionisation of
     ultra-cool atmospheres and explore the effect of environmental high-energy radiation on the atmosphere's degree of ionisation. We consider the effect of photoionisation by Lyman continuum
     radiation in three different environments: the InterStellar Radiation Field (ISRF), O and B stars
     in star forming regions, and  also for white dwarf companions in  binary
     systems.  We apply our Monte Carlo  radiation transfer code to investigate the effect of Lyman continuum
     photoionisation for prescribed atmosphere
     structures for very low-mass objects.}% We utilise atmosphere structures from the {\sc Drift-Phoenix}     model atmosphere grid.}
%Results: 
{The external radiation environment plays an important role for the
  atmospheric ionisation of very low-mass, ultra-cool objects. Lyman
  continuum irradiation greatly increases the level of ionisation in
  the uppermost atmospheric regions. Our results suggest that a shell
  of an almost fully ionised atmospheric gas emerges for brown dwarfs
  in star forming regions and brown dwarfs in white dwarf binary
  systems.  As a consequence, brown dwarf atmospheres can be
  magnetically coupled which is the presumption for chromospheric
  heating to occur and for Aurorae to emerge.  First tests for assumed
  chromosphere-like temperature values suggest that the resulting
  free-free X-ray luminosities are comparable with those observed from
  non-accreting brown dwarfs in star forming regions.}
{} %The environmental irradiation determines if the ultra-cool object will form an ionosphere (e.g. in the ISM), or if the magnetic coupling is strong enough as precursor for subsequent chromospheric heating (e.g. BD-WD binary).  }  {}


  \maketitle
  
\section{Introduction}

Radio and X-ray emission from ultra-cool objects has been reported by
\cite{Berger2002, 2004ApJ...615L.153S, Hallinan2006, Osten2009, Route2012, Burgasser2013, 2015AJ....150..180B, Williams2014, 2016ApJ...818...24K}  for example, and surveys are being conducted  (\citealt{Antonova2013,2016ApJ...826...73P,2017ApJ...846...75P}). The white-flare observations by \cite{Schmidt2016} (for ASASSN-16AE) and \cite{Gizis2013} (for W1906+40) provide evidence for stellar-type magnetic activity to occur also in brown dwarfs, and  hence, \cite{Schmidt2015} and
\cite{Sorahana2014} suggest that brown dwarfs have chromospheres.  \cite{2004ApJ...615L.153S} study the old brown dwarf G569Bab and conclude that coronal emission remains powerful also beyond  young ages. The presence of a magnetised atmospheric plasma, including a sufficient number of electrons, is required to allow the formation of a chromosphere/corona through MHD processes like Alfv{\'e}n wave heating similar to the Sun (\citealt{Mullan2016};
\citealt{Reep2016}; \citealt{Brady2016}). Seed electrons are also required to understand aurora on moon-less brown dwarfs.
%  \cite{2017ApJ...846...75P} provide a comprehensive summery of high-energy observations associated with magnetic activity in ultra-cool stars.

 \citet{Rodriguez-Barrera2015} show that thermal ionisation can
 produce a partially ionised gas in a substantial volume of cool brown
 dwarfs and giant gas planet atmospheres and a highly ionised gas in
 the hotter/younger brown dwarfs or M-dwarfs.  However, for low
 effective temperatures and low surface gravity, additional mechanisms
 are required to ionise the upper atmosphere to a degree that radio
 and X-ray observation become feasible as was also suggested by
 \cite{Mohanty2002}.  \cite{Rimmer2013} demonstrate that cosmic ray
 irradiation significantly enhances the electron fraction compared to
 thermal ionisation in the upper atmospheres of brown dwarfs, but the
 local degree of ionisation does not exceed $10^{-7}$ in these
 low-density regions.  \cite{2013P&SS...77..152H} and
 \cite{Rimmer2013} do further point out that CRs can affect the upper
 portion of the mineral clouds that form in brown dwarf atmospheres
 (\citealt{HellingCasewell2014}), an effect that is well established
 for solar system objects (\citealt{Leap1_2016}). The present paper
 takes the idea of environmental effects on the ionisation of
 atmospheres of very low-mass, ultra-cool objects one step further.

Ultra-cool objects, i.e. brown dwarfs and free-floating planets, are
observed in a large variety of environments.  Brown dwarfs and free
floating planets in star forming regions are exposed to a stronger
radiation field than objects that are situated in the interstellar
medium (ISM). Star forming regions host O and B stars that produce a
substantial fraction of high-energy radiation that may lead to the
ionisation of the outer atmosphere of brown
dwarfs. \cite{Sicilia-Aguilar2008} and \cite{Forbrich2007} show that
non-accreting brown dwarfs in star forming regions have X-ray
luminosities of the order of  log$_{10}$L$_{\rm
  x}$ $\approx 28$\,[erg\,s$^{-1}$] for
0.5\,\dots\,8 KeV.  \cite{Kashyap2008} present X-ray observations for
planet-host stars, including M-dwarfs and brown dwarfs
(M2\,\dots\,M8.5) with X-ray luminosities of log$_{10}$L$_{\rm
  x}$=26.36\,\dots\,31.22\,erg\,s$^{-1}$. This X-ray emission is
therefore likely to originate from chromospheric activity of the brown
dwarfs and low-mass stars.
% where the required pool of electrons could be enabled through external irradiation.

In this paper we investigate the effect of Lyman continuum radiation
(from photons with energies above the Lyman limit of E$>13.6$eV, hence
of wavlenegths $\lambda<912$\AA)  in different galactic environments
on the ionisation of the atmospheric gas of brown dwarfs. We study
three different cases: i) irradiation from the interstellar radiation
field (ISRF), ii) irradiation from O and B stars within a star forming
region, and iii) irradiation from a white dwarf companion.  Our
interest in case (iii) is supported by the recent finding of atomic
line emission from a brown dwarf in the WD-BD binary WD0137-349
(\citealt{2017MNRAS.471.1728L}).
%We apply a Monte Carlo photoionisation code to prescribed brown dwarf model atmospheres.  
We utilise the analysis frame work from \cite{Rodriguez-Barrera2015}
to study the resulting plasma parameters and the possible magnetic
coupling of the atmosphere.  Our results support the expectation that
brown dwarfs form chromospheres, and we make  the first suggestions for the
X-ray flux possibly emerging from a shell of optically thin, but hot
gas that may form the outer part of brown dwarf atmospheres. Our
approach is summarised in Sect.\,\ref{s:apps}.  We present our results
in Sec.\,\ref{s:results}. Section~\ref{discussion} contains our
discussion, and Sect.~\ref{s:con} contains our conclusions.

\section{Approach} \label{s:apps}

\begin{figure}
\centering
\hspace*{-0.9cm}\includegraphics[angle=0,width=0.6\textwidth]{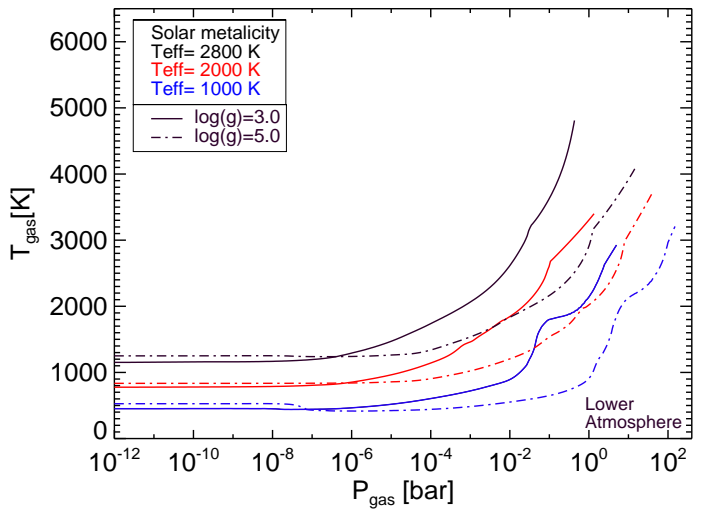}\\*[-0.8cm]
\hspace*{-1cm}
\caption{(T$_{\rm gas}$-p$_{\rm gas}$) profile from the {\sc
    Drift-Phoenix} model atmosphere simulations. (The (p$_{\rm
    gas}$-z) correlation can be found in Fig.~\ref{fig:Pgasz}.)  These models are used
  as input for the Monte Carlo photoionisation calculation. The
  hottest atmosphere represents a late M-dwarf or young brown dwarf
  atmospheres. The coolest atmosphere represents a planetary regime
  atmosphere. }
 \label{fig:Tpgas}
\end{figure}

We utilise {\sc Drift-phoenix} atmosphere model structures (T$_{\rm
  eff}$ = 2800\,K, 2000\,K, 1000\,K, $\log(g)$\,= 3.0, 5.0, {\small
  [M/H]}=0.0;
\citealt{Helling2008b,2008ApJ...677L.157H,Witte2009,Witte2011}) as
input for Monte Carlo Radiative transfer photoionisation
calculations. Using global parameters that describe young and old
brown dwarfs (T$_{\rm eff}$ [K], $\log(g)$ [cm\,s$^{-2}$], [M/H]), the
model atmosphere simulations {\sc Drift-phoenix} provides the local
gas properties (T$_{\rm gas} $[K], p$_{\rm gas}$ [bar], p$_{\rm e}$
[bar]; Fig.~\ref{fig:Tpgas}). {\sc Drift-phoenix} also calculates the
detailed cloud structure in ultra-cool atmospheres
(\citealt{Woitke2004,Helling2006,HellingCasewell2014}). For this
paper, we use the cloud particle number density, n$_{\rm d}$, (for
more on cloud details, see e.g. \citealt{2016MNRAS.460..855H}) to see
if Lyman continuum photons could reach deep enough into the atmosphere
to also be a source for cloud particle ionisation similar to cosmic
ray ionisation. This is, however, only a side-line of the present
paper.

We evaluate the effect of external Lyman continuum (LyC)  irradiation in three different environments: \\ i) {\it ISM:} This includes old brown dwarfs and free floating planes irradiated by the interstellar radiation field (ISRF). We use the ISRF  as lower limit for the Lyman continuum
irradiation causing possible atmospheric ionisation in brown
dwarfs.\\ ii) {\it Star forming region:} Brown dwarfs reside in
environments with strong radiation field, e.g. in star forming
regions with O and B stars.
 %the Lyman continuum irradiation from massive stars like an OB stars at a distance of 0.5 pc. 
 O stars allows us to explore an upper limit of the irradiation effect due to LyC, while B stars are more common members of star forming regions, like e.g. in the Orion OB1 stellar association.\\
%The effect of O and B stars is expected to be considerably larger than that from the ISM (see Table\,\ref{tab}). \\
 %{\it The main observable implication of the effect of Lyman continuum irradiation from OB stars on a brown dwarf atmosphere is that the probability of having a larger volume susceptible to emit at high energy levels (e.g at X-ray energies) has considerably increased}. \\
 iii) {\it White dwarf - brown dwarf binaries:} Old brown dwarfs can
 be companions of a white dwarf.  As an example, we utilise the binary system WD0837+185 composed of a cool white dwarf and an old brown dwarf with an orbital
 separation of d=0.006AU (\citealt{Casewell2015})\footnote{WD0837+185 is a binary system  composed of a white dwarf and a low mass companion
   with a close orbital separation ($<$\,3 AU). Such systems are very rare. The low number of such known systems might be linked to their
   formation mechanism  (\citealt{Casewell2013,
     Casewell2012}). WD0837+185 is one of the
   four white dwarf-brown dwarf binary systems known (\citealt{Casewell2015}). %Most of the substellar companions of Sun-like stars are planets and/or some brown dwarfs. It is not well understood how a low mass companion survives to the evolution of  its host star when it becomes a red giant (\citealt{Maxted2006}).
   }
   Despite a low occurrence rate of WD-BD binaries of only 0.5\% (\citealt{2013MNRAS.429.3492S}), these systems are an important link to giant gas planets since WD-BD pairs are far easier to observer.

\subsection{Photoionisation by Lyman continuum irradiation}\label{s:MCRT}

In an external radiation field originating from an O or a B star, a
white dwarf, or the ISRF, the radiation energy E\,$>$\,13.6 eV first
dissociates the atmospheric H$_{2}$ which leaves the gas in the upper
atmosphere to be dominated by atomic hydrogen.  This is supported by
\cite{Rimmer2015} (their Fig. 9) who show that H$_{2}\rightarrow$ H in
the uppermost atmospheric layers of an irradiated planet. 
  %This assumption may, however, weaken in the case of less strong
  %irradiation as for example for Jupiter with its H$_2$ dominated
  %atmosphere.  
  \cite{2008ApJ...677..790L} point out the Jupiter's
  upper atmosphere is dominated by H$_2$, that H$_3^+$ has been
  consistently predicted by models to be a major component of
  Jupiter's ionosphere, but that H$^+$ predominates at higher
  altitudes and on the night side. H$_3^+$ has a short life time,
  hence, decaying into atomic hydrogen quickly. But H$^+$ is long-lived as it
  required three-body electron recombinations or a radiative
  recombination reaction. \cite{2014IJAsB..13..173R} show that the
  dominating ionic molecule resulting from cosmic-ray triggered ion-neutral
  chemistry is NH$_4^+$, which also leads to H$_2$ dissociation. More
  work is required to solidify this argument also with respect to
  super-thermal electrons possibly emerging. However, we note that only a
  small fraction ($\sim$ 0.1\%) of the atmosphere will be ionsied such that the bulk ($\sim$ 99.9\%) 
  remains H$_2$-dominated. We now apply a Monte Carlo radiative
transfer (MCRT) to investigate the ionisation of atomic hydrogen by
Lyman continuum photons.  The abundance of H$^{+}$, hence, the number
of electrons originating from Lyman continuum irradiation, is derived
by applying a Monte Carlo photoionisation code (\citealt{Wood2000}).

% In this work the photoionisation simulation is set up to be a 1D plane parallel atmosphere. 
% Photons may therefore only exit the simulation via the upper and lower $z$ faces, thus simulating a semi-infinite plane parallel atmosphere. The $z$ direction traces the vertical 1D atmosphere profile where T$_{\rm gas}$, p$_{\rm gas}$, and $\rho_{\rm gas}$ increase inwards with increasing $z$ (Fig.\,\ref{fig:Tpgas}). 
%  By adopting a static plane parallel 1D atmosphere we determine the depth dependent ionisation structure in $z$-direction only. The $z$ component of this simulation is represented by results from 1D {\sc Drift-Phoenix} model atmosphere simulations (T$_{\rm gas}$(z), p$_{\rm gas}$(z), p$_{\rm e}$(z)).
The code balances photoionizations with radiative recombinations by the ionisation equilibrium equation in each cell to get a new ionization structure, %(changing the H$^{+}$ abundance in each cell),

\begin{equation}
n(H)\int _{ \nu_{o} }^{\infty }{ \frac {4\pi J_{\nu} }{ h\nu}} a_{\nu}(H)d\nu=n_{e}n_{p}\alpha(H,T)\,,
\end{equation}
where n$(H)$ is the neutral hydrogen atom (HI) density [cm$^{-3}$],
$J_{\nu}$ is the mean intensity [erg\,cm$^{-2}$\,s$^{-1}$\,Hz$^{-1}$],
a$_{\nu}(H)=6.3\cdot 10^{-18}$ cm$^2$ is the ionisation cross section
for the hydrogen with E $> h\nu_{\rm o}$ (E $>$ 13.6 eV),
$\alpha(H,T)$ is the recombination coefficient [cm$^{3}$\,s$^{-1}$],
and n$_{\rm e}$ and n$_{\rm p}$ [cm$^{-3}$] are the electron and proton
densities, respectively.  The opacity is updated according to the neutral
fraction of hydrogen and the next radiation transfer iteration is
carried out.  The Monte Carlo radiation transfer code tracks photon
paths and computes the ionisation structure of hydrogen in a three
dimensional grid. We use a 1D atmospheric density grid with input from  {\sc Drift-phoenix} models. It is homogeneously extended onto the 3D MC grid. Hence, we
utilised a 3D slab structure with an atmospheric density gradient in
order to see if a principle effect of LyC irradiation on ultra-cool
atmospheres occur. We are interested in the depth to which external
radiation may potentially ionise the atmosphere, in addition to cosmic
rays as demonstrated in \cite{Rimmer2013}.

\paragraph{Input parameters:}
The Monte Carlo code follows the random walks within the atmosphere
grid for both direct ionising photons from the source(s) and diffuse
photons produced by radiative recombinations direct to the ground
state of hydrogen. We adopt the same approach as \cite{Wood2000} and assign all direct stellar photons an energy of 18eV and diffuse photons an energy of 13.6eV. The hydrogen ionisation cross sections at these energies are $6.3\cdot 10^{-18}$ cm$^2$ and $2.7\cdot 10^{-18}$ cm$^2$ respectively. Our simulations only track the ionisation of hydrogen, so this two-energy approximation for the direct and diffuse photons is sufficient to determine the depth dependent ionisation structure. If we wanted to know the detailed depth-dependent ionisation structure of an element with multiple ionisation stages, then we would require an input spectrum as described in the Monte Carlo photoionisation code of \cite{2004MNRAS.348.1337W}.

 Direct ionising photons that are absorbed by
neutral hydrogen in the Monte Carlo simulation are converted to
diffuse photons with a probability $\alpha_{1}/ \alpha_A$, where
$\alpha_{1}$ is the hydrogen recombination coefficient direct to the
ground state and $\alpha_{\rm A}$ is the recombination coefficient to
all levels. We adopt a radiative recombination rate $\alpha_A = 4
\times 10^{-13}$~cm$^{3}$\,s$^{-1}$ and the ratio $\alpha_{1} /
\alpha_{A}$ = 0.38, which are appropriate for photoionised gas at
$10^4$~K (\citealt{Osterbrock2006}).  A key input parameter for the
Monte Carlo photoionisation simulations is the flux of LyC photons
at 18eV (688\AA) reaching the top of the atmospheres from the
different sources (Table\,\ref{t:tab}). It is determined by the adopted ionising luminosities and source-atmosphere distances.

 We conducted test calculations to determine how different input
 photon energies (corresponding to different cross section values) may
 affect our resulting values for the ionisation penetration depth.  We
 find that for the high densities in the brown dwarf atmospheres we
 are studying, that it is the total ionising luminosity that is the
 most important parameter for determining the depth of photoionisation
 and that the energy values adopted for the direct and diffuse photons
 have a negligible effect.

\begin{table}%\centering
%   \hspace{-0.3cm}
   \scriptsize\addtolength{\tabcolsep}{+0pt}
   \renewcommand{\arraystretch}{2}
    \fontsize{7}{6.2}\selectfont 
   \begin{tabular}{lllll}
    \hline\hline
  \multirow{2}{*}{Source}&
       \multicolumn{1}{c}{T$_{\rm eff}$} &
            \multicolumn{1}{c}{Q} &
      \multicolumn{1}{c}{d} &
       \multicolumn{1}{c}{$F_{\rm LyC}$}\\
     % \cline{2-5}
    &   \hspace*{+0.05cm} [K]& \hspace*{+0.1cm} [photons s$^{-1}$] &\hspace*{+0.2cm} &  [photons s$^{-1}$ cm$^{-2}$]  \\
    \hline\hline
        \centering
   
   % \hline
   \hspace*{+0.2cm} O3& 51230 &  \hspace*{+0.3cm}7.41$\times 0^{49}$ & \hspace*{+0.2cm}0.5 pc &  \hspace*{+0.5cm}2.5$\times 10^{13}$  \\
   % \hline
   \hspace*{+0.2cm}  B0 & 33340 &  \hspace*{+0.3cm}1.05$\times10^{48}$ & \hspace*{+0.2cm}0.5 pc &  \hspace*{+0.5cm}3.5$\times 10^{10}$  \\
   % \hline
   \hspace*{+0.1cm}  WD   & 14748 & \hspace*{+0.3cm}2.75$\times 10^{41}$ &  \hspace*{+0.1cm}0.006 AU&  \hspace*{+0.5cm}2.7$\times10^{18}$  \\
  
     \hspace*{+0.1cm} ISRF &\hspace*{+0.cm}--------  & \hspace*{+0.3cm}-----------   &\hspace*{+0.2cm}----------- &     \hspace*{+0.5cm}3.0$\times 10^{7}$ \\
       \hline
  \end{tabular}
       \caption{Ionising luminosities, $Q$, adopted for OB stars \citet{Sternberg2003}, white dwarfs (\citet{Casewell2012} and \citet{Hills1973}) and the resulting ionising fluxes, $F_{\rm LyC}$, reaching the top of the irradiated atmosphere for the quoted separations. For the ISRF, the ionizing flux is taken from \citet{Reynolds1984}.}
      \label{t:tab}
\end{table}

%The Lyman continuum photons  are emitted from O and B stars (e.g. \citealt{WoodLoeb2000}), white dwarfs (\cite{Casewell2012}; \cite{Hills1973}) or from the interstellar medium (e.g. \citealt{Sternberg2003}; \citealt{Reynolds1984}) that are emitted with energies above the Lyman limit. The ionising photons from the high-mass stars produce a free-free continuum fluxes (\citealt{Sternberg2003}).

\subsection{Free-free emission}
The ionised hydrogen gas in the upper atmospheres of irradiated brown dwarfs may emit
free-free radiation (thermal Bremsstrahlung) if the temperature is high enough. If the gas is optically thin, the resulting luminosity is calculated by
\begin{equation}
%\rm L^{\rm ff}_{\rm tot} (\rm z)=4\pi\, {(R+(z_{\rm max}-z_{\rm i}))}^2\,\epsilon_{\nu}^{\rm ff}\,\delta(\rm z),
\rm L^{\rm ff}_{\rm tot}=4\pi\, \int^{\rm z_{\rm max}}_{\rm R}{(R+(z_{\rm max}-z))}^2\,\epsilon_{\nu}^{\rm ff}(z)\,dz,
\label{eq:Lff}
\end{equation}
%\begin{equation}
%L^{\rm ff} (z)={(4\pi)}^2\, {(\Delta \rm {z})}^2\,\it {j}^{\rm ff},
%%\label{eq:Lff}
%\end{equation}
 with R [cm] the radius of the irradiated object, and z$_{\rm max}$ [cm] the maximum value of the vertical geometrical extension of the atmosphere. 
 %z$_{\rm i}$ [cm] the value of the vertical geometrical extension of a particular shell of the atmosphere, $\delta(\rm z)=\rm z_{\rm i+1}-\rm z_{\rm i}$ [cm] the vertical geometrical separation between two adjacent shells of the atmosphere.  
 The volume emissivity or the total free-free emission power that a gas can emit per unit of volume, per solid angle and per unit of frequency, $\epsilon_{\nu}^{\rm ff}(z)$ [erg\,cm$^{-3}$\,s$^{-1}$], of an optically thin gas integrated over a range of energy (\citealt{Osterbrock2006}) is given by the follow equation 
 \begin{equation}
\epsilon^{\rm ff}=5.44\cdot10^{-39}\,\rm T^{1/2}_{\rm em}\,\rm n_{\rm e}\rm n_{\rm i}\, \overline{g}_{B}\frac{1}{\rm h}\int_{\rm h\nu_{1}}^{\rm h\nu_{2}} e^{-\frac{h\nu}{\rm k_{\rm B}\rm T}}d(h\nu)\,. \label{eq:eff}
\end{equation}
%The resulting total luminosity from free-free emission  is L$^{\rm ff}_{\rm tot}=\sum_{\rm z} \rm
%L^{\rm ff}(\rm z)$ [erg s$^{-1}$].  
The electron density, $n_{\rm e}$ [cm$^{-3}$]
results from the Monte Carlo calculation.  T$_{\rm em}$ [K] is
the temperature of the emitting gas, $k_{\rm B}$ [eV\,K$^{-1}$] is the
Boltzamnn constant, $h$ [eV\,s] is the Planck constant and $\overline
\rm \overline{g}_{\rm B}$=1.2 is the Gaunt factor
%\footnote{The Gaunt
%  factor is a multiplicative factor that correct the calculations of
  %continuous absorption or emission when a classical method is used
  %rather than a quantum one.}  
  for a hydrogen plasma
(\citealt{Osterbrock2006}).
%The Gaunt factor results from the need to accurate free-free emission calculations over a wide range of plasma conditions (required for computational spectrum modelling). 
The integration boundaries h$\nu_{1}$=0.5 KeV and h$\nu_{2}$= 8 KeV correspond to X-ray range of the electromagnetic spectra. 
 
This simple approach allows us to test  first estimates  for the X-ray luminosity an
irradiated brown dwarf may emit. No model for the formation
of a brown dwarf chromosphere/corona is available to 
consistently derive the X-ray luminosity.  \cite{Mullan2016} suggest that Alfv{\'e}n waves are an efficient mechanism for heating the corona of M-dwarfs. They model a mechanical energy flux (Alfv{\'e}n waves) to explain observed X-ray fluxes from the corona of M-dwarfs. 
To the authors knowledge, only \cite{2013AN....334..137W} performed a
consistent modelling of a chromosphere for a M-dwarf. We therefore
follow an approach similar to  \cite{Feigelson2003, 2010A&A...511A..83F,
  Schmidt2015}. \cite{2005A&A...439.1137F} construct a semi-empirical chromosphere model  for quiescent  mid-M dwarfs following examples from the solar community (see references in their paper). 
\cite{2010A&A...511A..83F} apply this model to the flaring M5.5 star CN Leo. A linear temperature rise in transition region and chromosphere is modelled and  the gradient adjusted. The top of the chromosphere has a prescribed temperature of 8000K (quiescent chromosphere) or 8500K (flaring chromosphere). Higher temperature were considered in order to represent the heating by flares.  \cite{Schmidt2015} follow this approach by replacing the temperature in the outer atmosphere with a chromospheric temperature structure consisting of two components (their Fig. 8). The chromospheric temperature rise, the chromosphere break, and the start of the transition region  are treated as free parameters, given that not much is known about the formation of chromospheres on brown dwarfs and late-M-dwarfs.  They calculate activity strengths for their set of 11820 late-M- and brown dwarfs.
  Here, the 1D model consists of an underlying
(cloud-forming) {\sc Drift-Phoenix} atmosphere in radiative-convective
equilibrium and a temperature inversion region representing a chromospheric temperature increase.
 
\begin{figure*}
\centering
\includegraphics[angle=0,width=0.9\textwidth]{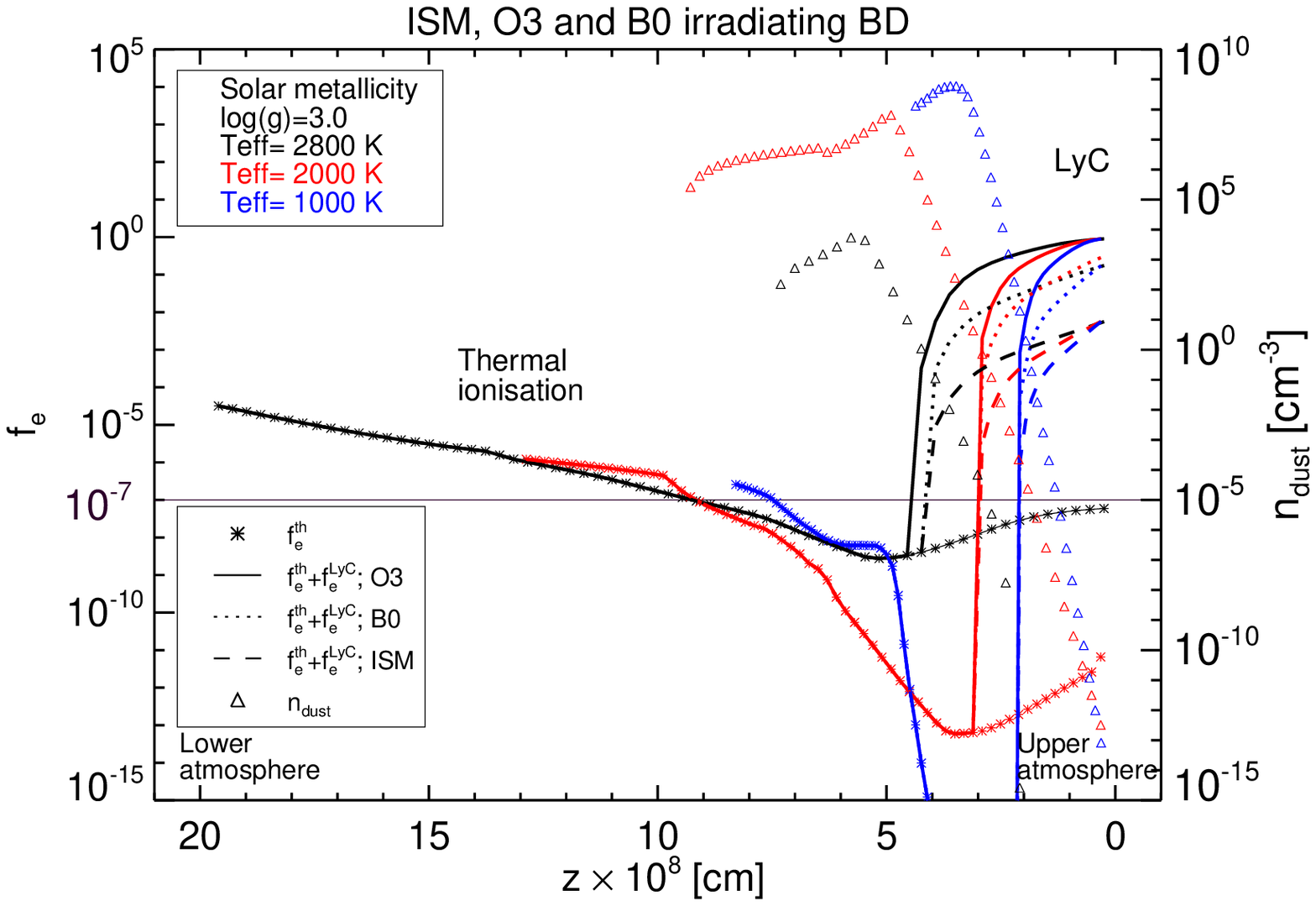}\\*[-0.6cm]
\includegraphics[angle=0,width=0.9\textwidth]{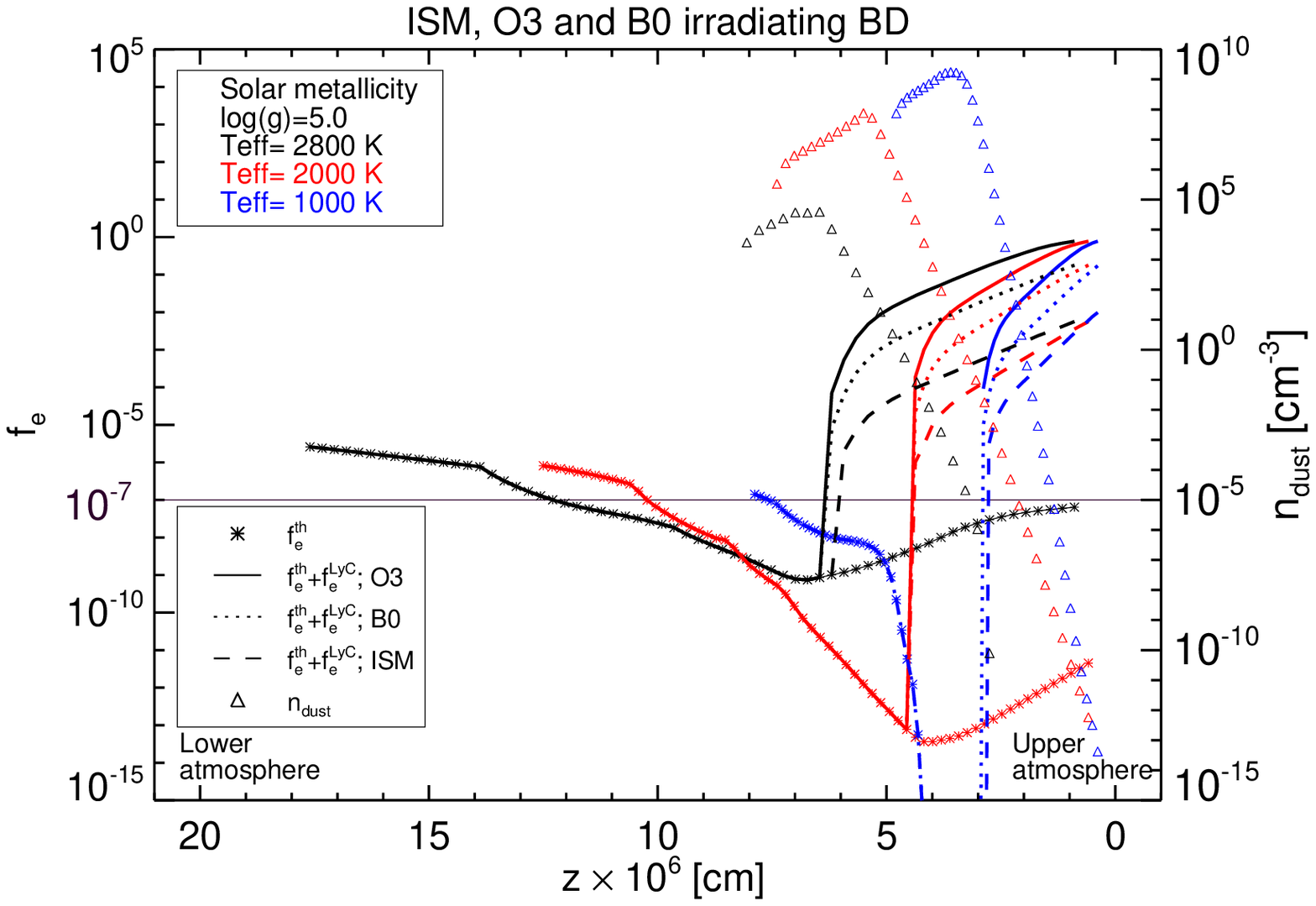}
\\*[-0.6cm]
\hspace*{+1.3cm}\caption{Effect of Lyman continuum irradiation from an O3 and a B0 star at a distance of 0.5 pc and from the ISRF compared to the effect of the thermal ionisation on  giant gas planet or young brown dwarf atmospheres ({\bf top}), and on  old brown dwarf atmospheres ({\bf bottom}). The threshold $f_{\rm e}>10^{-7}$ (thin solid black line) marks the point above which the atmospheric gas starts to be partially ionised. The cloud particle number density, n$_{\rm dust}$ [cm$^{-3}$], is plotted as comparison to where the cloud layer is located relative to the  LyC gas ionisation depth.}\label{f:fevsz}
\end{figure*}

\begin{figure*}
\centering
\includegraphics[angle=0,width=0.9\textwidth]{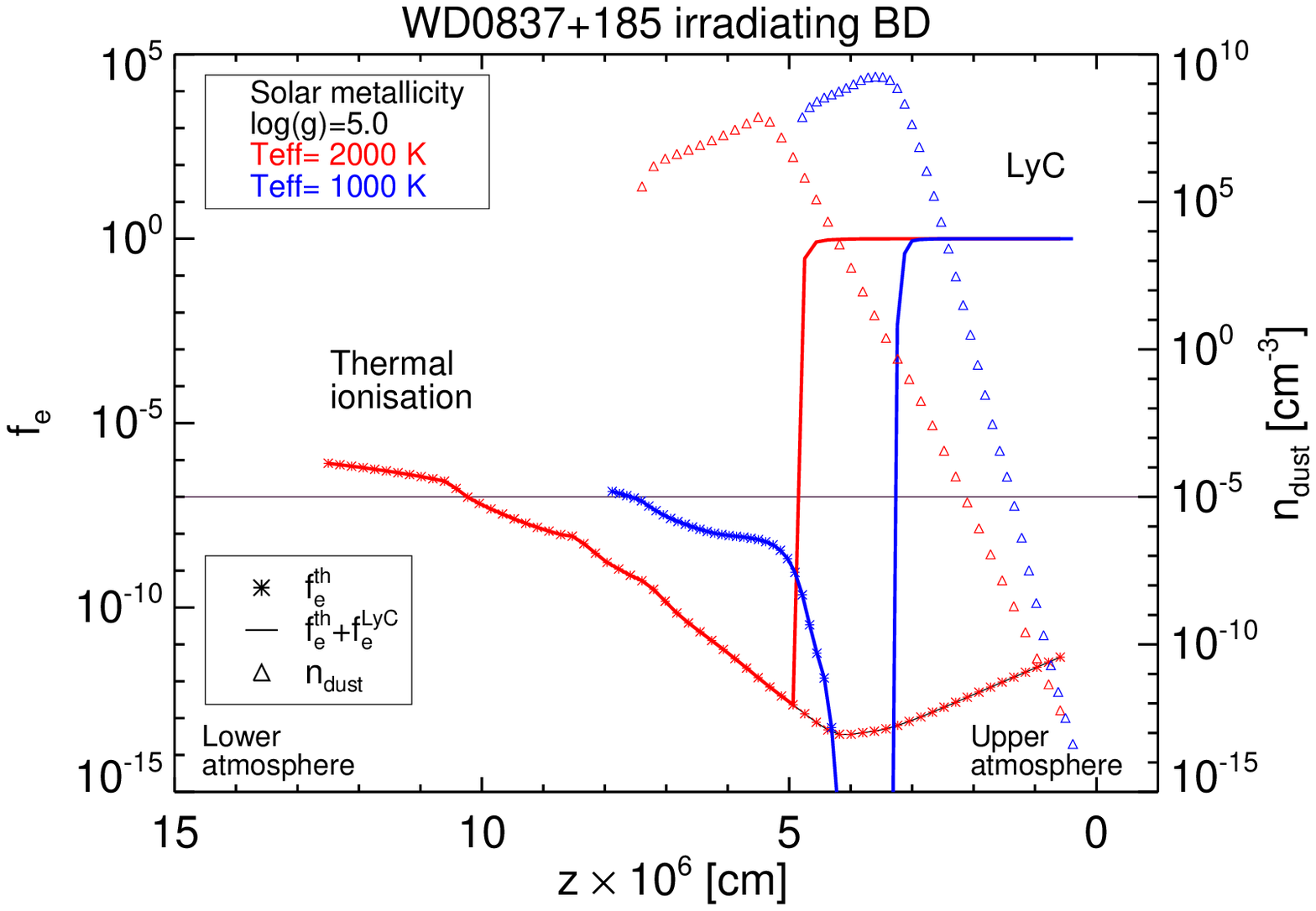}\\*[-0.6cm]
\vspace{+0.2cm}\caption{The effect of Lyman continuum irradiation  in comparison to  thermal ionisation in  old brown dwarf atmospheres for different T$_{\rm eff}$ in a binary system irradiated by a WD star at a distance of d=0.006 AU. Brown dwarf atmospheres can only be thermally ionised in deeper layers. Lyman continuum irradiation completely ionised the upper most layers.  The cloud particle number density, n$_{\rm dust}$ [cm$^{-3}$], is plotted as comparison to where the cloud layer is located relative to the LyC gas ionisation depth.
  } \label{f:fefpWD}
\end{figure*}

\section{Results} \label{s:results}

In the following, we evaluate the effect of Lyman continuum (LyC)
irradiation at 18eV on the ionisation in a brown dwarf atmosphere in
the three different scenarios i) -- iii) as introduced in
Sect.~\ref{s:apps}.  Case ii) and iii) are studied in detail with
respect to the effect of LyC irradiation on the electrostatic
character and regarding the potential magnetic coupling of the
atmosphere. We compare the effect of LyC to our previous reference
results for local thermal ionisation
(\citealt{Rodriguez-Barrera2015}). Case i) will only briefly be
summarised in Sect.~\ref{ss:ISRF} and then used as a lower-limit
reference for cases ii) and iii).  Our analysis is presented for each
plasma and magnetic parameter separately to allow a comparision
between the different irradiating environments
(Sects.~\ref{ss:ne}\,\ldots\,\ref{sss:mag}).

\subsection{Effect of the interstellar radiation field}\label{ss:ISRF}

 The interstellar radiation field (ISRF) present in the ISM is a lower
 limit for high-energy irradiation of an old brown dwarf without a
 companion (free floating objects).  It has a flux $\sim 10^{7}$
 photons\,s$^{-1}$\,cm$^{-2}$ (\citealt{Reynolds1984};
 Table\,\ref{t:tab}). The interstellar radiation field ionises the
 upper-most atmosphere layers of an ultracool objects (dashed lines in
 Fig.\,\ref{f:fevsz}) considerably more than thermal ionisation
 (asterisks).  The upper atmosphere layers reach a degree of
 ionisation $f_{\rm e}\sim 10^{-2}$ ($ f_{\rm e}=n_{\rm e}/(n_{\rm
   e}+n_{\rm gas})$, n$_{\rm gas}$ [cm$^{-3}$] -- gas density, n$_{\rm
   e}$ [cm$^{-3}$] -- local electron number density). The inwards
 increasing gas density decreases the influence of the LyC from the
 ISRF on the local degree of ionisation rapidly.

\subsection{The effect of Lyman continuum irradiation on the atmospheric electron budget}%\hspace{+3cm} Degree of ionisation, $f_{\rm e}$}
\label{ss:ne}

Figure\,\ref{f:fevsz} shows that the LyC irradiation reaches a similar
atmospheric depth, hence ionises a similar portion of the atmosphere
of a brown dwarf, independent of the incident LyC flux arriving from
different radiation sources (Table\,\ref{t:tab}). In the case of a
young brown dwarf (log(g)=3.0), the atmospheric portion ionised by
the LyC photons has a geometric extension of up to 5000km, for an older
brown dwarf (log(g)=5.0) it affects only the upper 50km as the
atmosphere is far more compact. These geometrical extensions reach a
P$_{\rm gas}=0.1\,\ldots\,10$bar in the coolest examples studied here
(T$_{\rm eff}=1000$K) but remain in the low-pressure regime for the
hottest atmosphere examples studied (T$_{\rm eff}=2800$K; see
Fig.~\ref{fig:Pgasz} for (P$_{\rm gas}$-z) correlation).  The degree to
which the gas is ionised will depend on the number of interacting
photons, hence on the environment where the brown dwarf resides in.

\paragraph{Star forming region (case ii): }\label{ss:starform_2}
The electron density from the photoionisation through the irradiation
of an O3 star is three orders of magnitude larger than that from the
ISRF in brown dwarf atmospheres with $\log(g)$=3.0 and two orders of
magnitude larger on brown dwarf atmospheres with
$\log(g)$=5.0. However, type B0 stars produce one and two orders of
magnitude more electrons than the effect of the irradiation from the
ISRF for brown dwarfs ($\log(g)$=\,3.0, 5.0).

Figure\,\ref{f:fevsz} shows that Lyman continuum irradiation provides
an efficient ionisation mechanism for the upper-most atmospheric
layers in ultra-cool objects.  The threshold $f_{\rm e}>10^{-7}$ marks
the point above which the atmospheric gas starts to be partially
ionised and hence, plasma behaviour may emerge
(\citealt{Rodriguez-Barrera2015}). In the case of O stars, a shell of
fully ionised gas forms in the uppermost atmosphere. This ionisation
effect cannot penetrate below a certain atmospheric depth, and other
mechanisms maybe needed to ionise the gas. For example, thermal ionisation
(asterisks) starts to dominate the ionisation of the atmosphere at
larger geometrical depth (i.e. higher pressures).

Figure \ref{f:fevsz} also depicts the location of the cloud
layer (open triangles) as a result of our {\sc
  Drift-Phoenix} atmosphere simulations. Only the outermost cloud layer
would be affected from the LyC, either by the high-energy photon
directly interacting with the cloud particles or from the
ionised gas depositing on the cloud particle surface. Both processes
are not considered here, but they can cause an ionisation of
the cloud particles similar to the electrification of the moon surface dust (\citealt{Leap1_2016}) as long as the cloud particles remain stable against
electrostatic disruption (\citealt{Stark2015, Helling2016}).

\paragraph{White dwarf - brown dwarf binary (case iii):}\label{ss:WD-BD_2}

Figure\,\ref{f:fefpWD} shows the degree of gas ionisation produced
from thermal ionisation and resulting from the white dwarf LyC flux
that impacts the brown dwarf atmosphere. Figure\,\ref{f:fefpWD}
suggests that the predominant ionisation process is Lyman continuum
irradiation in the upper 30-50km of an old brown dwarf's atmosphere,
and that it ionises the upper atmosphere completely for T$_{\rm
  eff}$=2000-1000\,K.  Also here, the uppermost layers of the cloud
where the smallest particles reside (haze layer; see
e.g. \citealt{2016MNRAS.460..855H}) will be affected by the LyC
irradiation.

\subsection{Plasma parameters}\label{sss:fp}

%The plasma frequency, $\omega_{\rm pe}$, measures the capacity of the plasma in response to any perturbation that can alter the electric field of the environment an is given by,

\begin{figure}
\centering
\hspace*{-1cm}\includegraphics[width=0.6\textwidth]{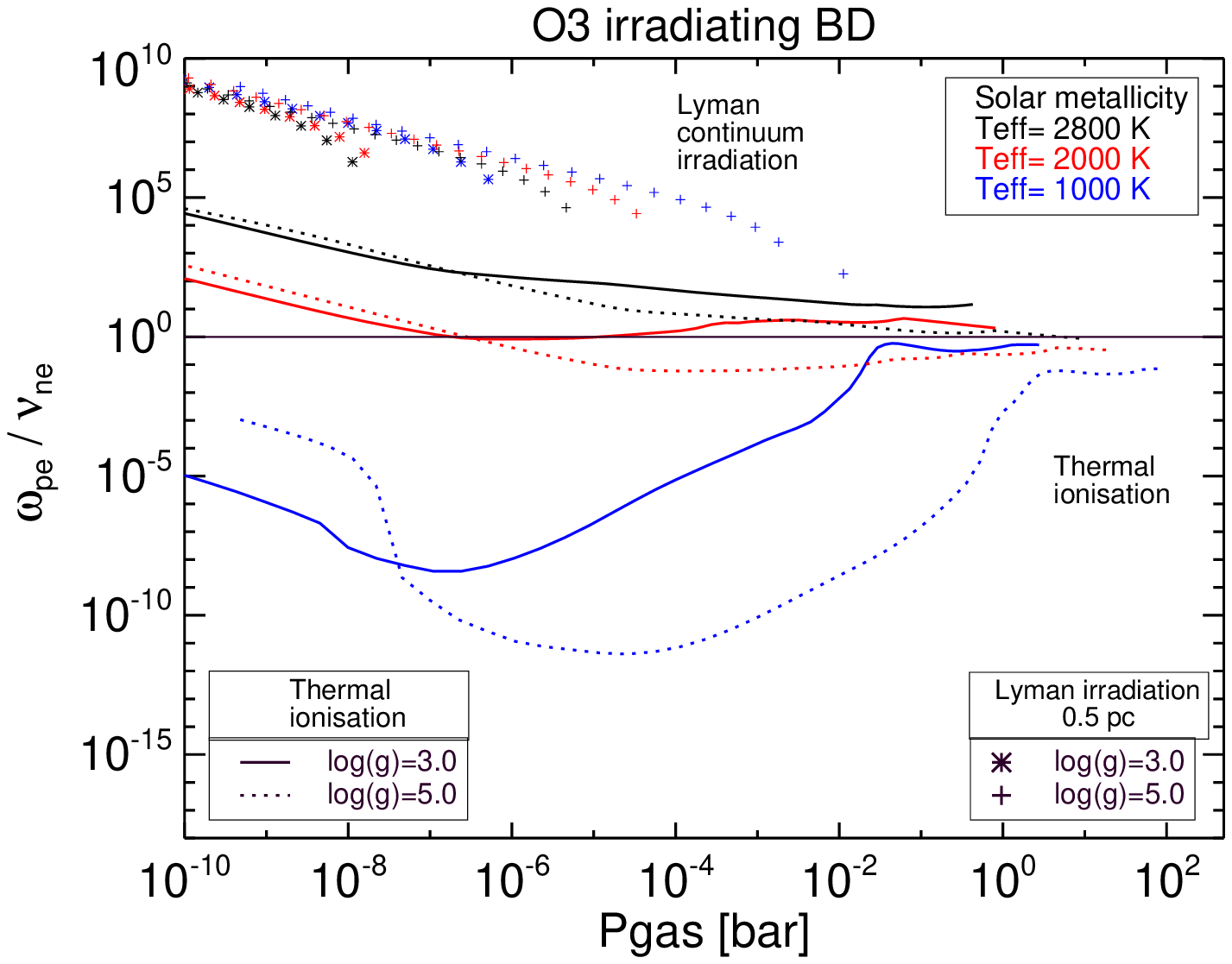}\\*[-0.6cm]
\hspace*{-1cm}\includegraphics[width=0.6\textwidth]{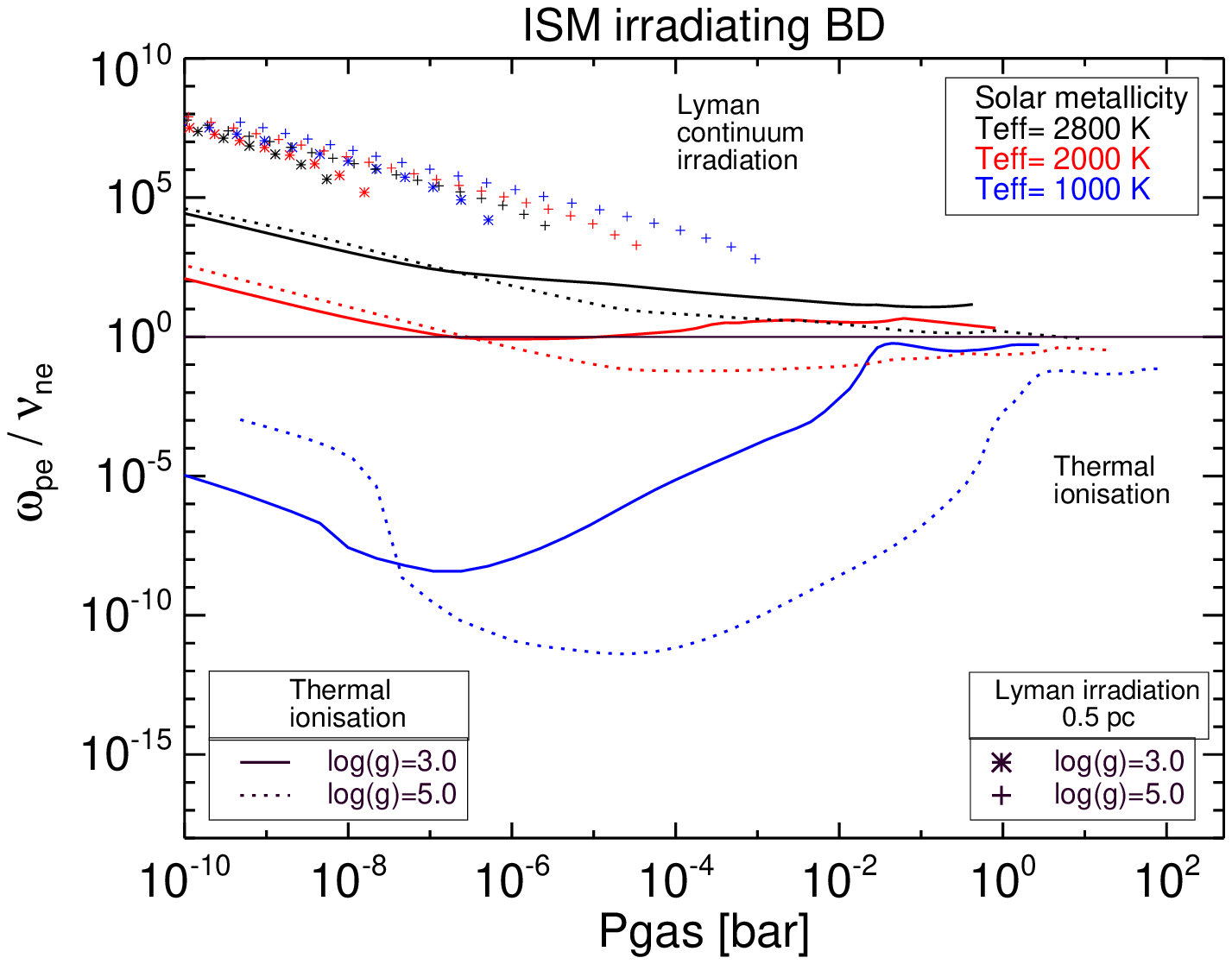}\\*[-0.6cm]
\hspace*{-1cm}\includegraphics[angle=0,width=0.6\textwidth]{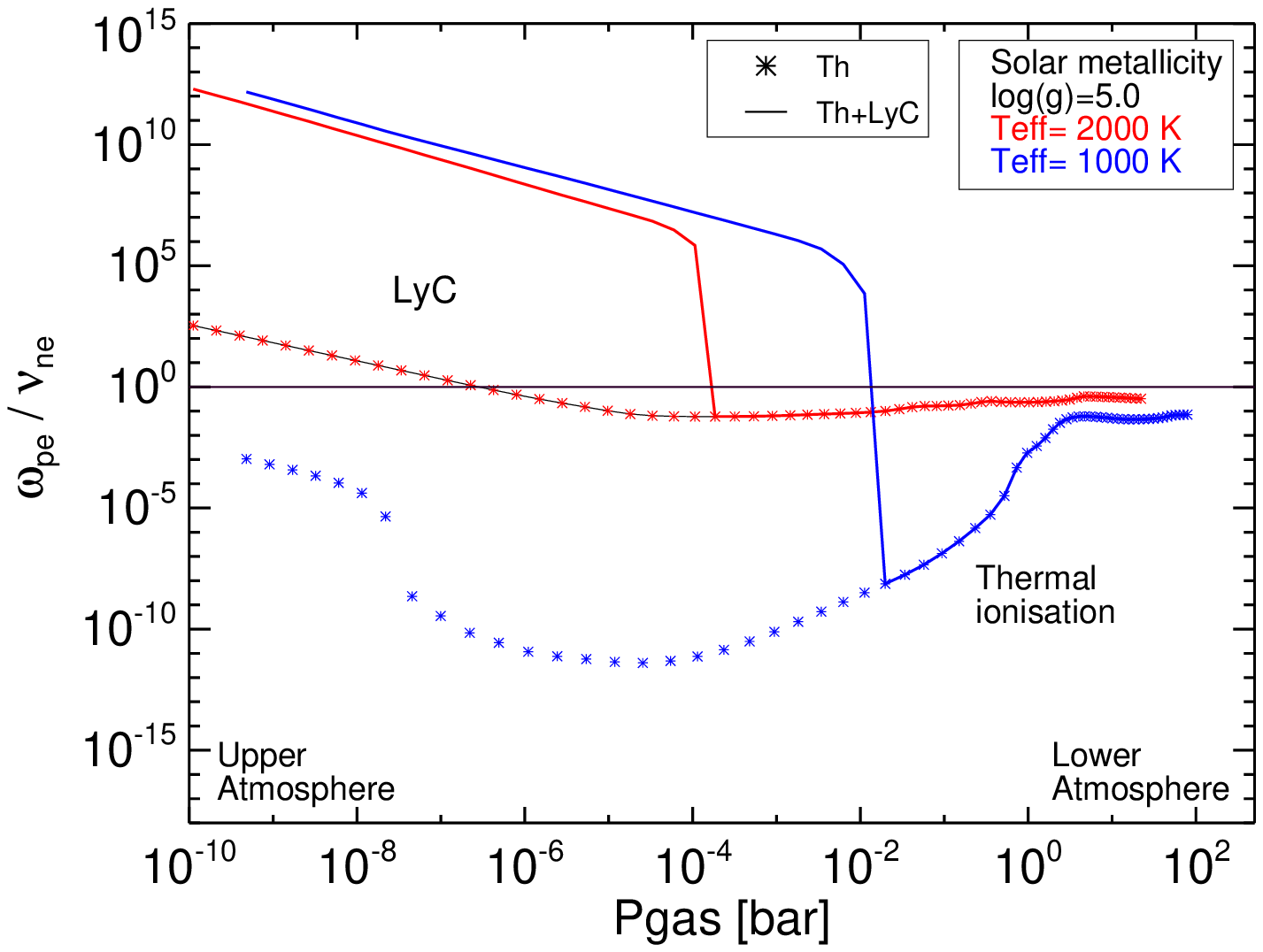}
\\*[-0.6cm]
\caption{The strength of electromagnetic interactions ($\omega_{\rm pe}/\nu_{\rm ne}$)  as result of  of LyC irradiation compared to thermal ionisation in brown dwarf atmospheres in different galactic environments. {\bf Top:}  (case ii) --  O3 star at a distance of 0.5 pc (star forming region), {\bf Middle:}  (case i) --  the ISRF, {\bf Bottom:} (case ii)  -- white dwarf - brown dwarf binary system  like WD0837+185 with d=0.006AU. Young brown dwarf atmospheres ($\log(g)$=3.0) and old brown dwarf atmospheres ($\log(g)$=5.0) are plotted.}\label{fig:wpeO3ISRF}
\end{figure}

If  $\omega_{\rm pe}\gg\nu_{\rm ne}$ ($\omega_{\rm pe}$ -- plasma frequency, $\nu_{\rm ne}$  -- collisional frequency neutrals-electrons; both in $[\rm s^{-1}]$),  than  electromagnetic interactions dominate over electron-neutral interactions in a gas. The plasma frequency is defined as
\begin{equation}\label{eq:fp}
\omega_{\rm pe}=\left(\frac{n_{\rm e}e^{2}}{\epsilon_{0}m_{\rm e}}\right)^{1/2} \,\,,
\end{equation}
with $e$ is the electron charge [C] and $m_{\rm e}$ is the electron mass [$\rm kg$], and 
the collision frequency for
neutral particles with electrons is,
\begin{equation}
 \nu_{\rm ne}=\sigma _{\rm gas}n_{\rm  gas}v_{\rm th,e } \,\,,
 \end{equation}
  where v$_{\rm th,e}$ is the thermal velocity of electrons ($v_{\rm th,e}$=$(k_{\rm B}T_{\rm s}/m_{\rm s })^{ 1/2 }$ [m\,s$^{-1}]$ ), and $\sigma_{\rm gas}$ is the scattering cross section of particles ($\sigma_{\rm gas}= \pi\,r_{\rm gas}^2=8.8\cdot10^{-17}$ cm$^{2}$; $r_{\rm gas}\approx r_{\rm H^{+}}=5.3\cdot10^{-9}$ cm).  
  Figure\, \ref{fig:wpeO3ISRF} show where in  the atmosphere the gas can be treated as a plasma. This study demonstrates that Lyman continuum irradiation increases the atmospheric volume where the collective long-range electromagnetic interactions dominate compared to our reference study (\citealt{Rodriguez-Barrera2015}).
    
  \paragraph{Star forming region (case ii):}\label{ss:starform_3}
  Figure\,\ref{fig:wpeO3ISRF} (top and middle panels) demonstrates
  that electromagnetic interactions dominate over collisional
  processes (i.e. $\omega_{\rm pe}/\nu_{\rm ne}\gg 1)$ for all models
  considered due to external LyC radiation. That is different to if
  only thermal ionisation (solid and dotted lines) is present.  The
  effect of external irradiation in  the form of LyC allows for a
  considerably larger atmospheric volume where electromagnetic
  interactions dominate compared to thermal ionisation
  alone. Figure\,\ref{fig:wpeO3ISRF} shows that different T$_{\rm
      eff}$ and $\log(g)$ result in at least one order of magnitude
    (or more) difference in the effect on the atmosphere. All
    atmosphere cases appear rather similar in the very top layers with
    respect to their capability for electromagnetic interactions.
 
\begin{figure}
\centering
\hspace*{-1cm}\includegraphics[angle=0,width=0.6\textwidth]{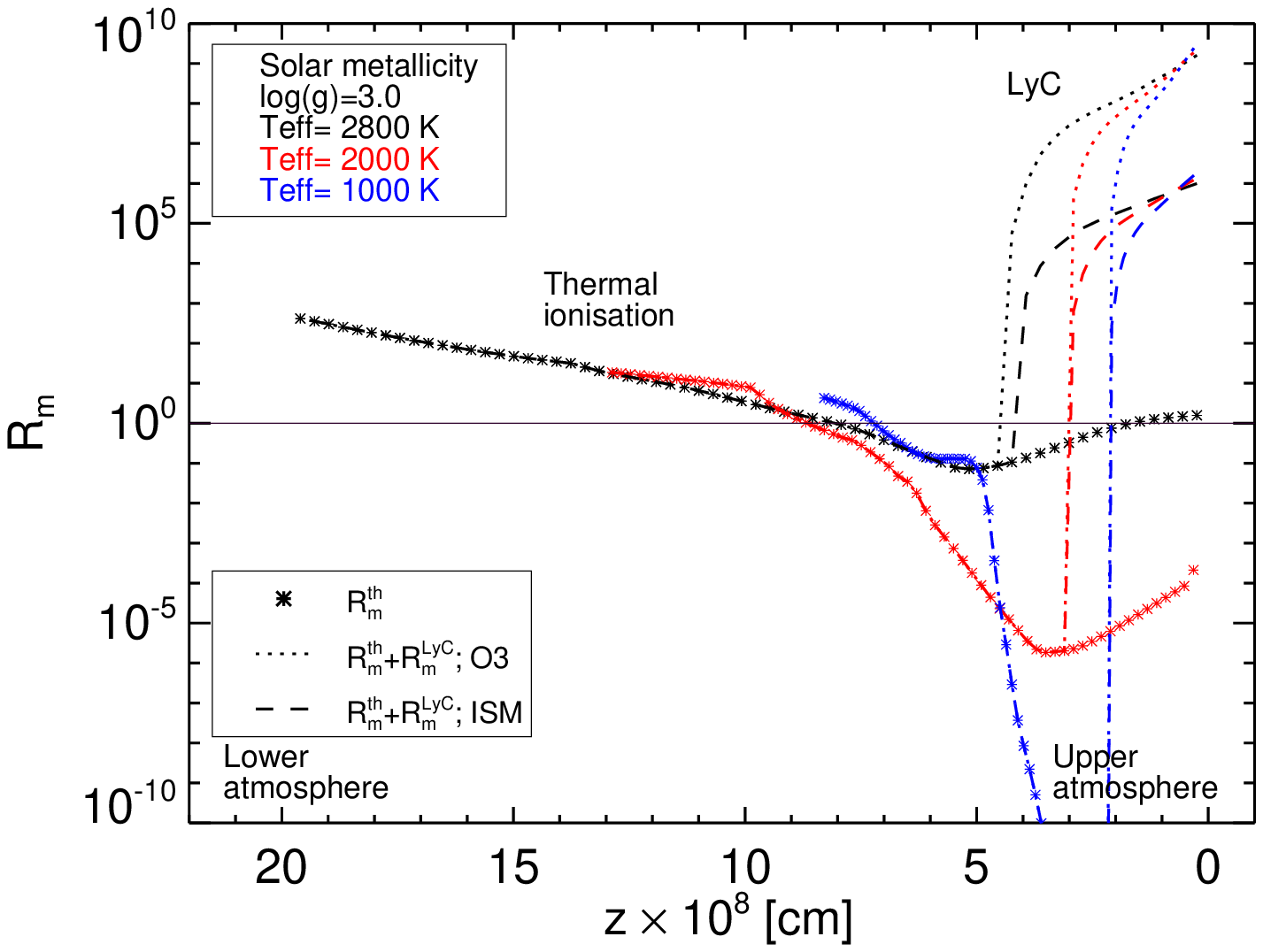}\\*[-0.6cm]
\hspace*{-1cm}\includegraphics[angle=0,width=0.6\textwidth]{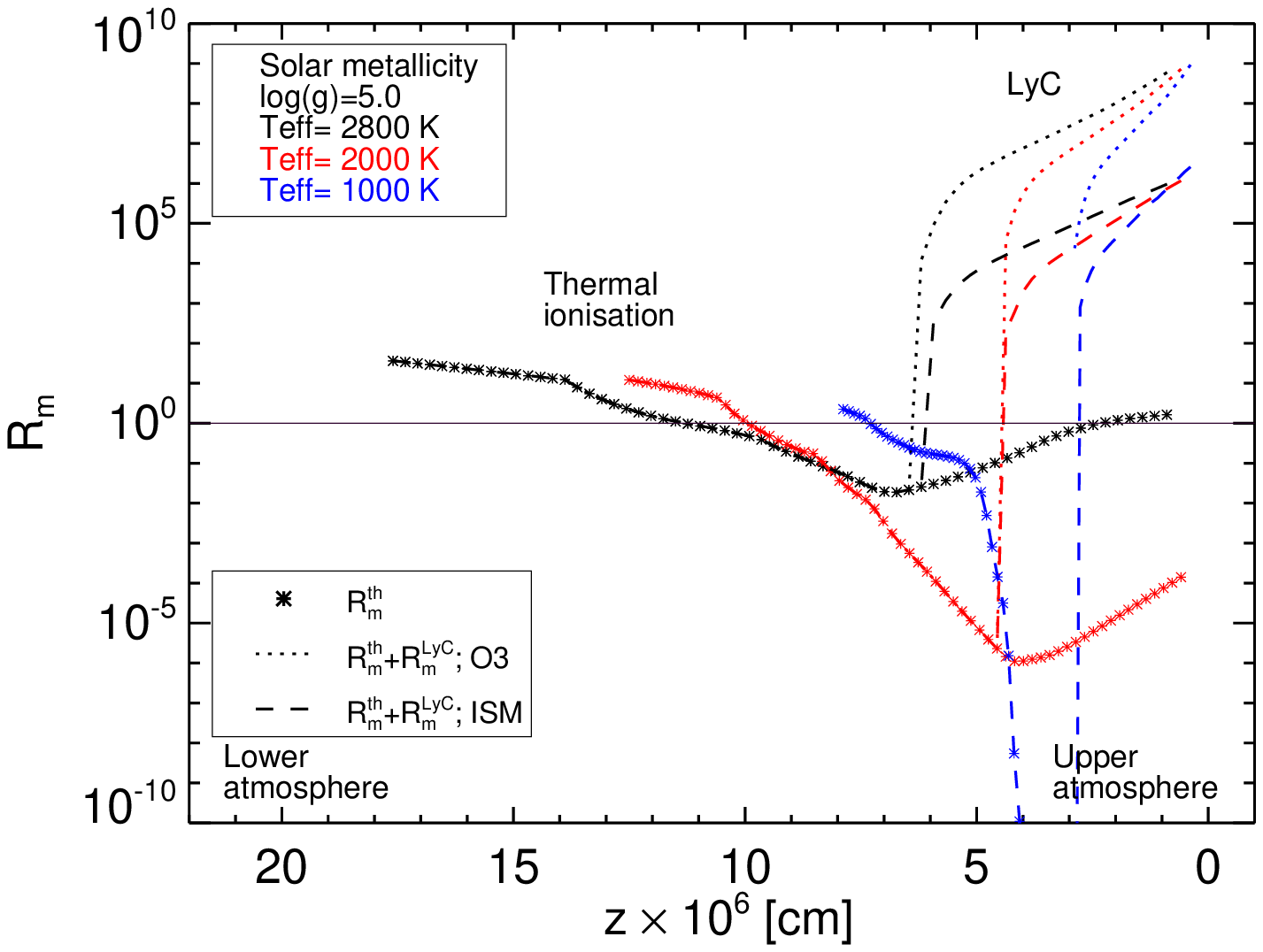}\\*[-0.6cm]
\hspace*{-1cm}\includegraphics[angle=0,width=0.6\textwidth]{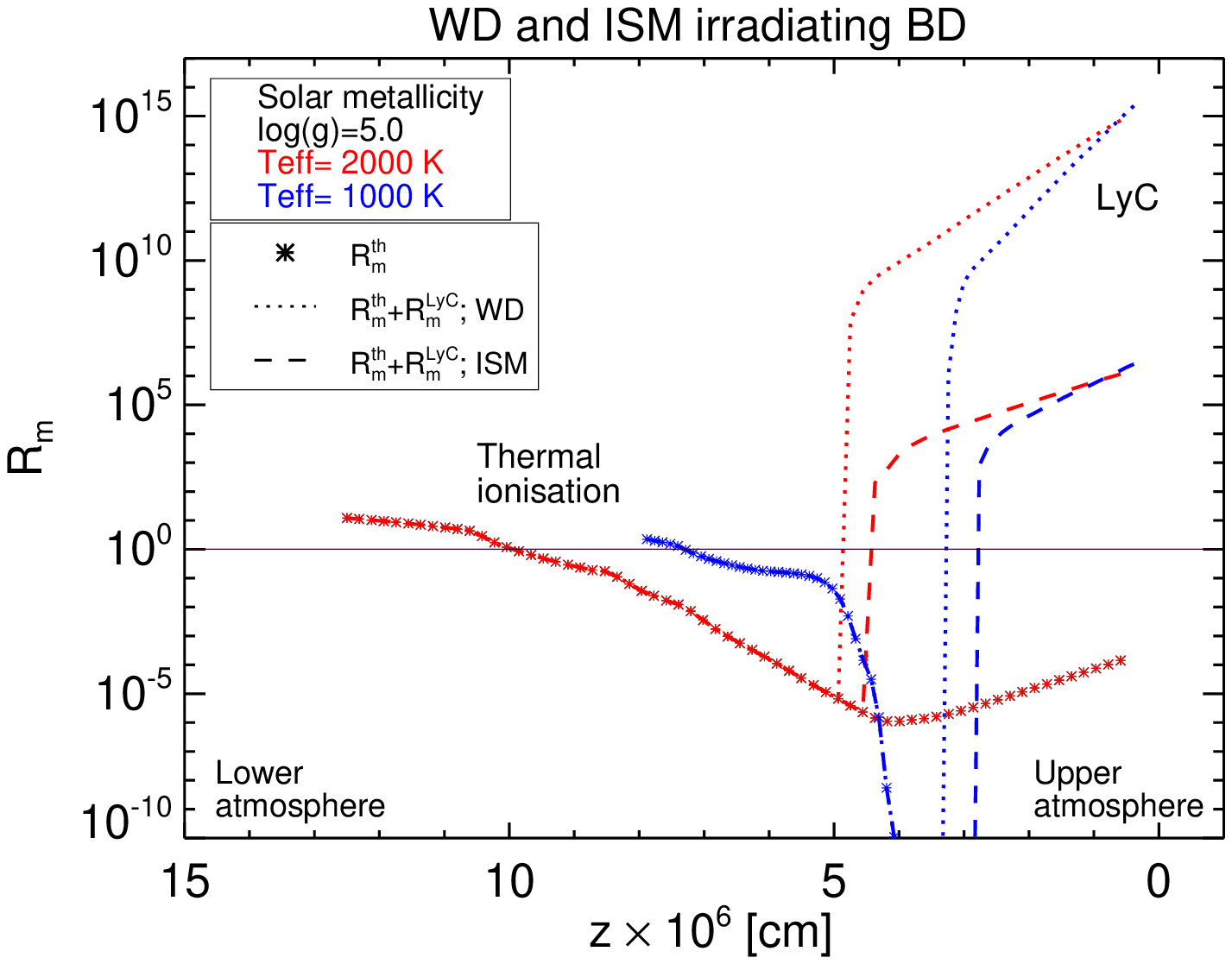}\\*[-0.6cm]
\caption{The magnetic Reynolds number, R$_{\rm m}$ ($v_{\rm flow}=10^{6}$ cm\,s$^{-1}$) as result of  LyC irradiation compared to thermal ionisation in different galactic environments: {\bf Top:}  young brown dwarfs in star forming region (O3 star, d=0.5pc), {\bf Middle:}  old brown dwarf in forming region (O3 star, d=0.5pc), {\bf Bottom:} Old brown dwarf in a white dwarf binary system like WD0837+185 with d=0.006AU. }
 \label{fig:Rmz}
\end{figure}
\paragraph{White dwarf - brown dwarf binary (case iii):}\label{ss:WD-BD_3}

Figure\,\ref{fig:wpeO3ISRF} (bottom panel) demonstrates that the collective long-range electromagnetic interactions dominate over the collisions between electron and neutral particles at p$_{\rm gas}\le$\,10$^{-4}$\, bar for the atmosphere with T$_{\rm eff}$=2000\,K and at p$_{\rm gas}\le$\,10$^{-2}$\, bar for the coldest atmosphere with T$_{\rm eff}$=1000\,K. The comparison of those results with the effect of the thermal ionisation shows that Lyman continuum irradiation increases the atmospheric volume where the collective long-range electromagnetic interactions dominate by six orders of magnitude in pressure, hence, the affected region increases substantially. The LyC irradiation from a white dwarf causes a larger volume of the brown dwarf atmosphere to form a plasma and it causes a stronger  electromagnetic interaction compared to LyC from O and B stars in  star forming regions (case ii) for a given atmosphere structure (compare panels in Fig.\,\ref{fig:wpeO3ISRF}).

\subsection{Magnetic parameters}\label{sss:mag}

In previous sections, we have quantified the degree of ionisation and
the plasma frequency and how they are affected by the Lyman continuum
irradiation in comparison to thermal ionisation in a substellar
atmosphere for different galactic environments. In this section, we
show if and how the magnetic coupling changes due to external LyC
photoionisation. The atmospheres are analysed with respect to a
critical magnetic flux and the classical Reynolds number.

A gas is magnetised if $\omega_{\rm c,s}\gg \nu_{\rm n,s}$
($\omega_{\rm c,s}$ -- cyclotron frequency, $\nu_{\rm n,s}$ --
collisional frequency, both [s$^{-1}$]). A critical magnetic flux,
$B_{\rm s}$, can be derived that is required to assure $\omega_{\rm
  c,s}/\nu_{\rm n,s}\gg 1$ (Sect 5.1 in
\citealt{Rodriguez-Barrera2015}):
\begin{equation}\label{eq:B}
 B_{\rm s}=\frac {m_{\rm s}}{e} \sigma_{\rm gas,e }n_{\rm gas }{ \left( \frac {k_{\rm B}T_{\rm s}}{m_{\rm s }} \right)  }^{ 1/2 } [T]\,\,,
\end{equation}

 where $\sigma_{\rm gas}= \pi\cdot r_{\rm gas}^2$ [m$^{2}$] is the
 collision, or scattering, cross section, m$_{\rm s}$ is the mass of
 the species $s$ in [kg] and T$_{\rm s}$ the temperature of the
 species in [K]. T$_{\rm s}$ = T$_{\rm gas}$ is assumed. Hence, if a
 local magnetic field $B\gg B_{\rm s}$, then magnetic coupling will be
 possible.

The magnetic Reynolds number, R$_{\rm m}$, provides a measurement of
the diffusivity of the magnetic field in a given atmospheric gas. The
magnetic Reynolds number can be calculated by R$_{\rm m}=vL/\eta$,
where $L$ [m] is the pressure scale length of the considered plasma
and $\eta$ $[\rm m^{2}\,s^{-1}]$ is the diffusion coefficient.  The
pressure length scale is assumed to be L=10$^{3}$ m being a
representative scale height value for a brown dwarf with $\log(g)=5$
(Helling et al. 2011). The diffusion coefficient, $\eta$, can be
approximated by $\eta\approx\eta_{\rm d}$
(\citealt{Rodriguez-Barrera2015}) being $\eta_{\rm d}$ the decoupled
diffusion coefficient $\eta_{\rm d}={c^{2}\nu_{\rm ne}}/{\omega_{\rm
    pe}^{2}} $.  The plasma has reached the ideal MHD regime if
R$_{\rm m}\gg1$ and the coupling between the plasma and the background
magnetic field is complete.
     \begin{figure*}
\centering
\hspace*{-0.9cm}\includegraphics[angle=0,width=0.9\textwidth]{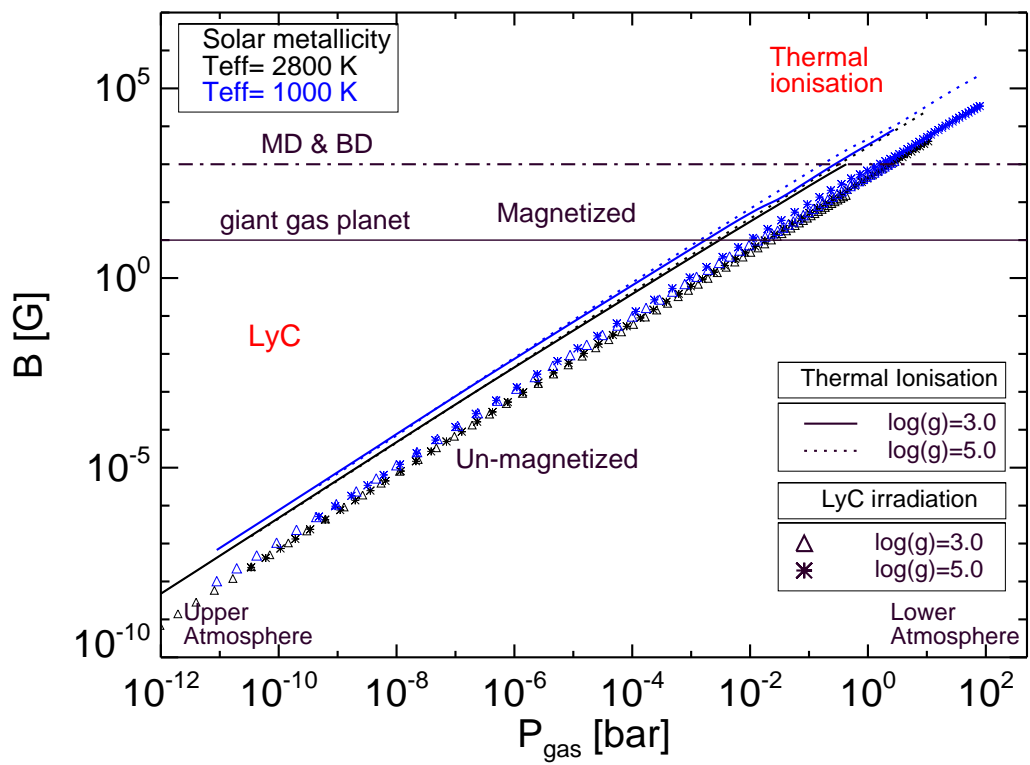}
\vspace{-0.3cm}\caption{The magnetic flux density, $B=B_{\rm s}$, required for the electrons to achieve magnetic coupling in ultra-cool objects. The effect of Lyman continuum irradiation ({\bf Asterisks} --  late M-dwarfs and young brown dwarf atmospheres, $\log(g)$=3.0; {\bf Triangle} --  old brown dwarfs, $\log(g)$=5.0) is compared to the effect of the thermal ionisation ({\bf Solid line} -- $\log(g)$=3.0; {\bf Dot line} --  $\log(g)$=5.0). Brown dwarfs are irradiated from an O3 star at a distance of 0.5 pc in a star forming region. 
 }
\label{fig:B}
\end{figure*}

\paragraph{Star forming region (case ii):}\label{ss:starform_4}

\vspace{+0.1cm} Figure\,\ref{fig:B} shows where electrons resulting
from LyC irradiation from an O star can be magnetised in a
M-dwarf/brown dwarf/planet atmosphere compared to thermal
ionisation, hence, where $B\gg B_{\rm s}$. We furthermore observe that
\begin{description}
 \item - for atmospheres with T$_{\rm eff}$=2800\,K (M-dwarfs and young brown dwarfs) a typical  background magnetic field flux of B$=10^{3}$ G magnetises the ambient electrons at p$_{\rm gas}< 5$ bar,
 \item - for atmospheres with T$_{\rm eff}$=1000\,K (planets), a typical background magnetic field flux of B$=10 $ G magnetises the electrons at p$_{\rm gas}< 5\cdot10^{-2}$ bar.
\end{description}

Figure\,\ref{fig:Rmz} shows the magnetic Reynolds number as the result of the thermal ionisation plus the photoionisation by Lyman continuum irradiation from an O star and from the ISRF. 
 The results from Fig.\,\ref{fig:Rmz} suggest that
\begin{description}
\item - a decrease of T$_{\rm eff}$, for a given $\log(g)$ and [M/H], decreases the atmospheric volume  where R$_{\rm m}\gg$1, 
\item - an ideal MHD behaviour (R$_{\rm m}\gg$1) can be considered in the upper atmosphere due to the significant increase of the electron density (R$_{\rm m} \propto n_{\rm e}$) by the effect of the Lyman continuum irradiation compared to the effect of thermal ionisation only. 
\end{description} 

\paragraph{White dwarf - brown dwarf binary (case iii):}\label{ss:WD-BD_4}

\begin{figure*}
\centering
\hspace*{-0.6cm}\includegraphics[width=0.9\textwidth]{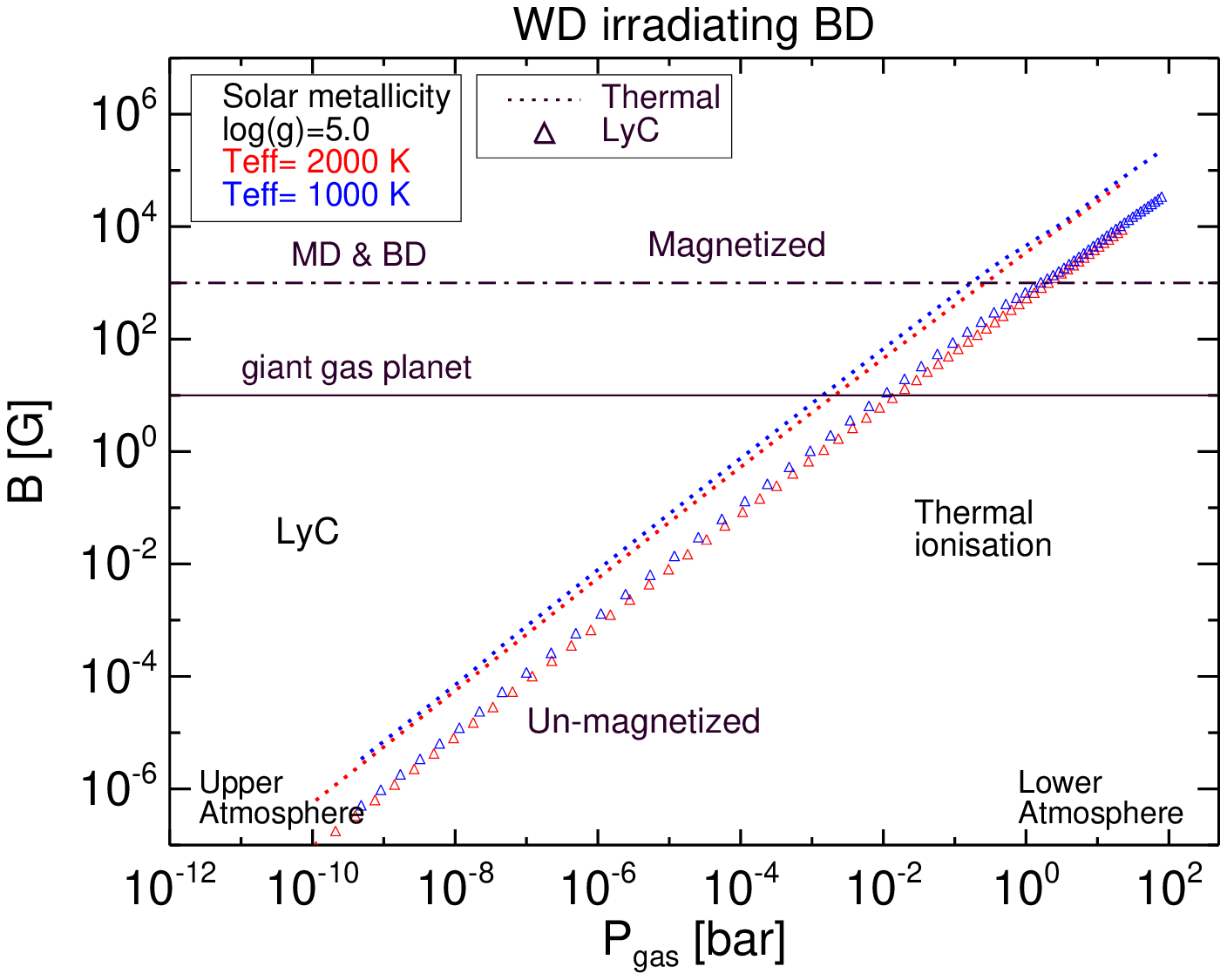}
\vspace{-0.5cm}\caption{Magnetic flux density required for the electrons to achieve magnetic coupling in ultra-cool objects. The effect of Lyman continuum irradiation is compared to the effect of the thermal ionisation. We consider a binary system with a WD star irradiating a brown dwarf at a distance of d=0.006 AU. }
\label{fig:BWD}
\end{figure*}

\vspace{+0.1cm}
Figure\,\ref{fig:BWD} demonstrates that LyC irradiation from white dwarfs would not only magnetise electrons in brown dwarf atmospheres but also in  M-dwarf  and giant gas planet atmospheres:
\begin{description}
 \item - for atmospheres with T$_{\rm eff}$=2000\,K (brown dwarfs, M-dwarfs) with a  background magnetic field flux of  B$=10^{3}$ G, electrons are  magnetised  for  p$_{\rm gas}<5$ bar,
  \item - for atmospheres with T$_{\rm eff}$=1000\,K (giant gas planets),  a  background magnetic field flux of  B=10\,G  can magnetise the electrons at p$_{\rm gas}< 5\cdot10^{-2}$ bar.
\end{description}

Figure\,\ref{fig:Rmz} (bottom panel) shows the magnetic Reynolds number for LyC irradiation from  a WD and the ISM compared to the effect of thermal ionisation. The results are similar to case (ii) except that the Reynolds number reaches higher values if the brown dwarf would be affected by LyC from a white dwarf.

\subsection{The free-free emission luminosity from strongly ionised, optical thin  atmospheres}\label{sss:L}

If we assume that these uppermost highly ionised atmosphere layers are
optically thin, we can estimate the resulting luminosity from
free-free emission (Eq.\,\ref{eq:Lff}) for the three cases of galactic
environment considered here. The emissivity $\epsilon^{\rm ff}\sim
T_{\rm em}^{1/2}n_{\rm e}n_{\rm i}\times f(e^{1/T_{\rm em}})$
(Eq.~\ref{eq:eff}) requires information about the local densities,
$n_{\rm e}, n_{\rm i}$, and the temperature of the emitting gas. The
local electron number densities are a direct result of our
calculation. For the local electron temperature, $T_{\rm em}$ [K], we
test two cases: $T_{\rm em}=T_{\rm gas}$ with $ T_{\rm gas}$ the local
LTE gas temperature as result of the {\sc Drift-Phoenix} model
atmosphere.  This case assumes that no effective heating of the
ionised gas occurs.  In the second case, we use $T_{\rm em}=T_{\rm
  chrom}$ with $T_{\rm chrom}=const$ a preset temperature of expected
chromospheric values, representing potential heating through
photoionisation and mechanical waves. We have no information about
chromospheric temperatures on brown dwarfs and suggestions in the
literature vary: \cite{Burgasser2013} suggest that the brown dwarf,
2MASS J13153094-2649513 (L7), emits electrons with temperature of
$T_{\rm em}\,\approx10^{9}-10^{10}$\,K, \cite{Osten2009} quote a
temperature of $T_{\rm em}\,\approx10^{6}$K, and \cite{Schmidt2015}
apply $T_{\rm em}=10^{4}$K. In order to provide first estimates based
on our ionisation calculations, we follow the examples by
\cite{2010A&A...511A..83F} and \cite{Schmidt2015} in postulating an
isothermal temperature layer with temperatures higher than in the
inner atmosphere. Such a layer could also be the result of MHD wave
heating as result of the magnetic coupling demonstrated in
Figs.~\ref{fig:B} and \ref{fig:BWD}. The temperature of the emitting
electrons may also decouple from the gas due to their significantly
lower mass relative to neutrals and ions. Hence, because the 
electrons will lose a smaller fraction of their energy during a
collision, they may not be in thermal equilibrium with their
surrounding in those regions where X-ray and radio emission emerge
from.  We test two values, $T_{\rm em}=10^5$K and $T_{\rm em}=10^6$K.

Table~\,\ref{Ltot} summarises the free-free emission luminosities,
L$^{\rm ff}_{\rm tot}$, that would occur if the LyC ionised gas were
optically thin and had the temperature $T_{\rm em}$. The highest
  amount of free-free emission luminosity is to be expected from brown
  dwarfs orbiting white dwarfs, the lowest from brown dwarfs in the
  interstellar medium. The strong increase of the values in
  Table~\,\ref{Ltot} when increasing the local gas temperature to
  $10^6$K results from the steepness of the high-energy tail of the Planck function.
 
 \begin{table*}
\centering
\caption{Upper limits for  X-ray luminosity  for 0.5\,\dots\,8 KeV from free-free emission, L$^{\rm ff}_{\rm tot}$ [erg\,s$^{-1}$], for optically thin, ionized gases. We assume ultra-cool objects  (T$_{\rm eff }$=2800\,K, 2000\,K, 1000\,K; $\log(g)$=3.0, 5.0) in case (ii) irradiated by an O3 or B0 star at distance d= 0.5 pc, in case (iii) irradiated by  a WD star at distance d=0.006 AU, and in case (i)  from the ISRF.  We neglect any shadowing effects. The temperature of the emitting gas is set to T$_{\rm em}=$10$^{5}$\,K and T$_{\rm em}$=10$^{6}$\,K to mimic a  chromospheric temperature or an otherwise thermally decoupled emitting electron population.}
% {\large L$^{\rm ff}_{\rm tot}$ [erg\,s$^{-1}$]} \vspace{+0.15cm} \\
 \hspace{-0.3cm}
   \scriptsize\addtolength{\tabcolsep}{+4.5pt}
   \renewcommand{\arraystretch}{2}
    \fontsize{7.5}{6.2}\selectfont
  \begin{tabular}{l|l|l|l|l|l}
    \hline\hline
     \multirow{2}{*}{Source } &
    \multirow{2}{*}{ \hspace{+0.1cm} T$_{\rm eff}$ [K]  }&
            \multicolumn{2}{c|}{ \hspace{-0.1cm}L$^{\rm ff}_{\rm tot}$ [erg\,s$^{-1}$] for T$_{\rm em}$=10$^{5}$\,K} &
      \multicolumn{2}{c}{ \hspace{+0.1cm}L$^{\rm ff}_{\rm tot}$ [erg\,s$^{-1}$] for T$_{\rm em}$=10$^{6}$\,K} \\
    %  \cline{3-8}
    &&     \hspace{+0.2cm}  $\log(g)$=3.0  & \hspace{+0.2cm}$\log(g)$=5.0  &   \hspace{+0.2cm}     $\log(g)$=3.0 & \hspace{+0.2cm}$\log(g)$=5.0 \\
    \hline
  %  O3 & & & & & & \\
    \hline
    \hspace{+0.1cm}  O3& \hspace{+0.3cm} 2800   &  \hspace{+0.3cm} 1.8$\cdot 10^{+2}$ &   \hspace{+0.3cm} 8.5$\cdot 10^{+1}$ &  \hspace{+0.3cm}7.0$\cdot 10^{+27}$  &  \hspace{+0.3cm} 8.6$\cdot 10^{+26}$  \\
   % \hline
   & \hspace{+0.3cm} 2000  &   \hspace{+0.4cm}3.4$\cdot 10^{+3}$&  \hspace{+0.3cm} 7.4$\cdot 10^{+3}$ &  \hspace{+0.3cm}3.4$\cdot 10^{+28}$  &  \hspace{+0.3cm} 7.5$\cdot 10^{+28}$ \\
   
& \hspace{+0.3cm} 1000  & \hspace{+0.4cm}1.6$\cdot 10^{+6}$  &  \hspace{+0.3cm} 1.2$\cdot 10^{+7}$ &  \hspace{+0.3cm}1.6$\cdot 10^{+31}$  &  \hspace{+0.3cm} 1.2$\cdot 10^{+32}$ \\

 \hspace{+0.1cm}  B0    & \hspace{+0.3cm} 2800   &  \hspace{+0.6cm}2.1 &   \hspace{+0.6cm}1.2  &  \hspace{+0.3cm}2.1$\cdot 10^{+25}$  &  \hspace{+0.3cm} 1.2$\cdot 10^{+25}$  \\
   % \hline
   & \hspace{+0.3cm} 2000  &    \hspace{+0.4cm}3.5$\cdot 10^{+1}$  &  \hspace{+0.3cm} 1.0$\cdot 10^{+2}$ &  \hspace{+0.3cm}3.6$\cdot 10^{+26}$ &  \hspace{+0.3cm} 1.0$\cdot 10^{+27}$ \\
   
& \hspace{+0.3cm} 1000  &  \hspace{+0.4cm}5.7$\cdot 10^{+3}$  &  \hspace{+0.3cm} 1.6$\cdot 10^{+5}$ &  \hspace{+0.3cm}5.7$\cdot 10^{+28}$   &  \hspace{+0.3cm} 1.6$\cdot 10^{+30}$ \\

    \hspace{0.0cm}  ISM    & \hspace{+0.3cm} 2800  &    \hspace{+0.4cm}1.8$\cdot 10^{-3}$  &   \hspace{+0.4cm}1.0$\cdot 10^{-3}$   &  \hspace{+0.3cm}1.8$\cdot 10^{+22}$   &  \hspace{+0.3cm} 1.0$\cdot 10^{+22}$   \\
   % \hline
   & \hspace{+0.3cm} 2000  &  \hspace{+0.4cm}1.0$\cdot 10^{-2}$  &  \hspace{+0.3cm} 8.9$\cdot 10^{-2}$ &  \hspace{+0.3cm}1.0$\cdot 10^{+23}$ &  \hspace{+0.3cm} 9.0$\cdot 10^{+23}$ \\
   
& \hspace{+0.3cm} 1000  &    \hspace{+0.6cm}5.2  &  \hspace{+0.4cm}1.4$\cdot 10^{+2}$&  \hspace{+0.3cm}5.3$\cdot 10^{+25}$  &  \hspace{+0.3cm} 4.0$\cdot 10^{+32}$ \\
     
      \hspace{+0.0cm}  WD & \hspace{+0.3cm} 2000  &\hspace{+0.5cm} ------ &   \hspace{+0.4cm}8.3$\cdot 10^{+12}$&\hspace{+0.5cm} ------ &   \hspace{+0.3cm} 8.4$\cdot 10^{+37}$ \\
   
& \hspace{+0.3cm} 1000    &\hspace{+0.5cm} ------ &    \hspace{+0.3cm} 2.5$\cdot 10^{+16}$ &\hspace{+0.5cm} ------&    \hspace{+0.3cm} 2.6$\cdot 10^{+41}$ \\

    \hline

  \end{tabular}\label{Ltot}
  
\end{table*}

 \paragraph{Star forming region (case ii):}\label{ss:starform_5}
  The free-free emission luminosities in the X-ray energy interval
  $0.5\,\ldots\,8$KeV in Table~\ref{Ltot} reach values
  comparable to observations from non-accreting brown dwarfs at X-ray
  wavelengths (\citealt{Sicilia-Aguilar2008}; \citealt{Kashyap2008};
  \citealt{Forbrich2007}; \citealt{Feigelson2003}) if the upper LyC
  ionised atmosphere can be heated substantially.  The emissivity is
  very low if the emitting part of the atmosphere would emit with the
  local LTE gas temperature
  (Fig.\,\ref{fig:Tpgas}). \citet{Schmidt2015} present the
  H$_{\alpha}$ emission from of ultra cool (M7-L8) brown dwarfs and
  suggest that these observations require the presence of a
  chromosphere.  \cite{Forbrich2007} suggest that the candidate brown
  dwarf B185839.6-365823, with a log\,L$_{\rm x}$=28.43 (L$_{\rm
      x}\sim2.7\cdot10^{28}$ erg\,s$^{-1}$) requires a high plasma
  temperature in order to fit the observed spectrum.  Another
    example for a very late dwarf, the M8.5 brown dwarf
    B185831.1-370456, was observed with a luminosity from free-free
    emission of log\,L$_{\rm x}$=26.9 (L$_{\rm x}\sim8\cdot10^{26}$
    erg\,s$^{-1}$).  \cite{Neuhauser1999} report a $\sim$\,1\,Myr old
  brown dwarf (Cha\,H$_{\alpha}$\,1) with a quiescent emission at
  X-ray wavelengths of log\,L$_{\rm x}$=28 [erg\,s$^{-1}$].  They
  suggest that magnetic activity must be present to explain those
  observations. \cite{Feigelson2003} detect 525 objects in a massive
  cluster (Orion nebula) at energies of 0.5\,\dots\,8 KeV
  (X-rays); 144 of those objects observed are cataloged as brown
  dwarfs (logM$\le$-0.2 M$\odot$) (their Table\,1) with an average of
  the X-ray emitted luminosity of log\,Lx $\approx$ 28.84
  erg\,s$^{-1}$. Our estimates of the free--free luminosity being
  emitted from brown dwarfs that form a shell of highly-ionised but
  optically thin gas in star forming regions are of the same order of
  magnitude. Our estimates, which are summarised in Table~\ref{Ltot},
  show that the resulting luminosity from free-free emission in the
  energy interval 0.5\,\dots\,8 KeV increases if:
\begin{description}
\item - the object forms a chromosphere with a substantially increased temperature compared to LTE values.
\item - the T$_{\rm eff}$ decreases.
%\item - the surface gravity increases.
\end{description}

An increase in surface gravity, hence a decrease in atmospheric scale heights, does lead to an increased luminosity in most but not all cases.

\paragraph{White dwarf - brown dwarf binary (case iii):}\label{ss:WD-BD_5}

Table~\ref{Ltot} also contains the results of the resulting luminosity
from free-free emission (thermal Bremsstrahlung irradiation) of a
brown dwarf that is ionised by Lyman continuum irradiation from a WD
star.  The results shows that L$^{\rm ff}_{\rm tot}$ is larger in a
white-dwarf-brown-dwarf system than in a star forming region or
through the ISRF due to the close orbital separation between the white
dwarf and the brown dwarf. Recent optical and near-infrared
  observations of the close white dwarf-brown dwarf non-interacting
  binary system WD0137-349 by \cite{2017MNRAS.471.1728L} show He, Na,
  Mg, Si, K, Ca, Ti and Fe lines being emitted from the brown dwarf
  companion. This is a strong indication for a temperature inversion
  in the upper atmosphere, and in combination with the detected
  H$\alpha$ emission a strong indication for a hight-temperature
  plasma being present in this cool compaion to a white dwarf. This
  conclusions for such systems is supported by
  \cite{2018MNRAS.tmp..241C} who detected Mg~I  and Ca~II emission lines in
  addition to  H$\alpha$ lines in the (shortest-period) non-interacting,
  white dwarf-brown dwarf post-common-envelope binary known EPIC
  21223532.

%\clearpage

\section{Discussion}\label{discussion}

Unexpectedly powerful emissions at radio, X-ray and H$_{\alpha}$ wavelengths from
very low-mass objects (ultra-cool dwarfs) have been observed by
different groups (e.g. \citealt{Williams2014}; \citealt{Burgasser2013}; \citet{Berger2002}; \citealt{Route2012}; see \citealt{2017ApJ...846...75P} for a recent survey).  
%A magnetised plasma leading to the presence of a chromosphere/corona seem required and must be sustained. In addition,  an efficient dynamo mechanism must exist to generate a large-scale magnetic field in such ultra-cool objects \citep{Cook2014}. 
%  \citet{Williams2014} show that ultra-cool objects do
%not follow the classical G\"udel-Benz relationship where the radio
%luminosity increases proportional to the X-ray luminosity in F\,$-$\,M
%stars \citep{Gudel1993}. 
%The deviation of ultra-cool stars in the
%G\"udel-Benz relationship beyond than approximately M5 may suggest a
%change in the dynamo mechanism that produces the magnetic field in
%such ultra-cool objects \citep{Cook2014}. 
\citet{Sorahana2014} suggest that weakened H$_2$O
(2.7$\mu$m), CH$_4$ (3.3$\mu$m) and CO (4.6$\mu$m) absorption in
combination with moderate H$\alpha$ emission could be linked to
chromospheric activity and its effect on the underlaying atmosphere structure.  Observations by \citet{Schmidt2016,Schmidt2015}  support this interpretation. They  postulate the presence of a chromosphere to reproduce extensive H$\alpha$-activity survey data for brown dwarfs.  \cite{Schmidt2016} report an old L0 dwarf emitting powerful emissions at H$_{\alpha}$ and near IR wavelengths. Such powerful emissions suggest that the magnetic activity must be present even in old type brown dwarfs and it implies the presence of a magnetised atmospheric plasma even in such ultra-cool objects. However, no consistent MHD simulations for brown dwarfs have been carried out yet. \cite{2013AN....334..137W}  studied the formation of a chromosphere on fully ionised M-dwarfs. Non-ideal (i.e. partially ionised) MHD simulations by 
\citet{2015ApJ...809..125T}  suggest the formation of a temperature inversion by magneto-convection processes  and Alfv\'en wave heating  (see also \citealt{Mullan2016}; \citealt{Reep2016}; \citealt{Brady2016}) in giant gas planets, a process that according to our work presented here could also work for brown dwarfs leading to the formation of a  chromosphere.
%The associated mass loss discussed by  \citet{2015ApJ...809..125T} would need to be re-addressed for brown dwarfs because brown dwarfs have a  surface gravity that can be two orders of magnitude larger than for giant gas planets. 
 
Another strand of addressing the radio emission from brown dwarfs are works like e.g.
 \cite{Nichols2012} who study the properties of radio emissions in ultra-cool dwarfs assuming the occurrence of auroral regions.   They suggest that the coupling between the atmospheric plasma and the magnetic field  effects a (postulated) high-latitude ionosphere  and generate auroral processes. A loss of particles to the ionosphere is suggested.
\cite{Spiers2014} present a theoretical
approach for cyclotron radio emission from Earth's auroral region showing that the radio emission results from a backward-wave cyclotron-maser emission process.

 \cite{2017ApJ...846...75P} have provided a concise overview of these different approaches suggesting that a transition from Sun-like chromospheric processes to Jupiter-like large-scale magnetospheric current systems occurs at the low-mass end of the main sequence. Our paper has added to this discussion by addressing the question of the origin of the pool of electrons that allow a chromosphere to form or a magnetospheric current system to develop, assuming that a magnetic field is present.  Although we do not present a consistent simulation to demonstrate the formation of a  chromospheric region on brown dwarfs,  
first estimates of the  X-ray luminosity of a hot, optically thin gas that is ionised by  Lyman continuum radiation suggest values that are comparable to observations of non-accreting brown dwarfs in star forming regions.

The required high electron density could affect the cloud layer in
brown dwarf atmospheres. \cite{Helling2016} suggest that cloud
particles can be destroyed if the electron temperature is $>
10^{5}$\,K. This would occur if a chromosphere-like temperature
increase would coincide with some part of the cloud layer where a
strong gas ionisation occurs (only the uppermost atmosphere,
Fig.\,\ref{f:fevsz}).

 \section{Conclusions}\label{s:con}

Lyman continuum irradiation causes a considerable increase of ionisation in the upper and outermost atmospheric regions of ultra-cool objects, forming a shell of substantial local ionisation. We demonstrated that these atmospheric regions exhibit a far stronger plasma than with thermal ionisation alone, and it would therefore be reasonable to call this part of the atmosphere an ionosphere. If a  sufficient global magnetic field is present  magnetic  coupling of these highly ionised atmospheric layers occurs, and we may therefore expect the formation of a chromosphere  if the object is sufficiently convective  such that  MHD heating mechanisms similar to M-dwarfs can occur (\citealt{2013AN....334..137W}).
 
Different galactic environments were investigated (ISRF, star forming region,  white dwarf - brown dwarf binary). Lyman continuum radiation from the interstellar radiation field has the smallest effect on the degree of  ionisation. The outer atmosphere of a brown dwarf binary as companion of a white dwarf, however, can be expected to be fully ionised.  More rigorous follow-up simulations are therefore warranted in order to study effects like, e.g., magnetically driven mass loss or even mass transfer in WD-BD systems.

%-------------------------------------------------------------------
\begin{acknowledgements}
M.I.R-B. and ChH highlight financial support of the European Community
under the FP7 by the ERC starting grant 257431. ChH highlights the
hospitality of the Kaptyn Astronomical Insititut at the University of
Groningen, and travel support from NWO and LKBF. We thank
A.~Sicilia-Aguilar and P.~Rimmer for the insightful and valuable
discussions of the manuscript. Most literature search was performed
using the ADS. We acknowledge our local computer support highly.
\end{acknowledgements}

\footnotesize{
\bibliographystyle{aa}
\bibliography{bib4}{}

%\newpage
%
\appendix
%\section{Summary of approaches used to represent chromospheres in M-dwarfs and brown dwarfs}

\section{Supplimentary details}

This appendix contains supplementary information about the relation between the local gas pressure, P$_{\rm gas}$ and the geometrical extension of the atmospheres, z in Figs.~\ref{fig:Pgasz}.

 \begin{figure}
\centering
\hspace*{-1cm}\includegraphics[angle=0,width=0.6\textwidth]{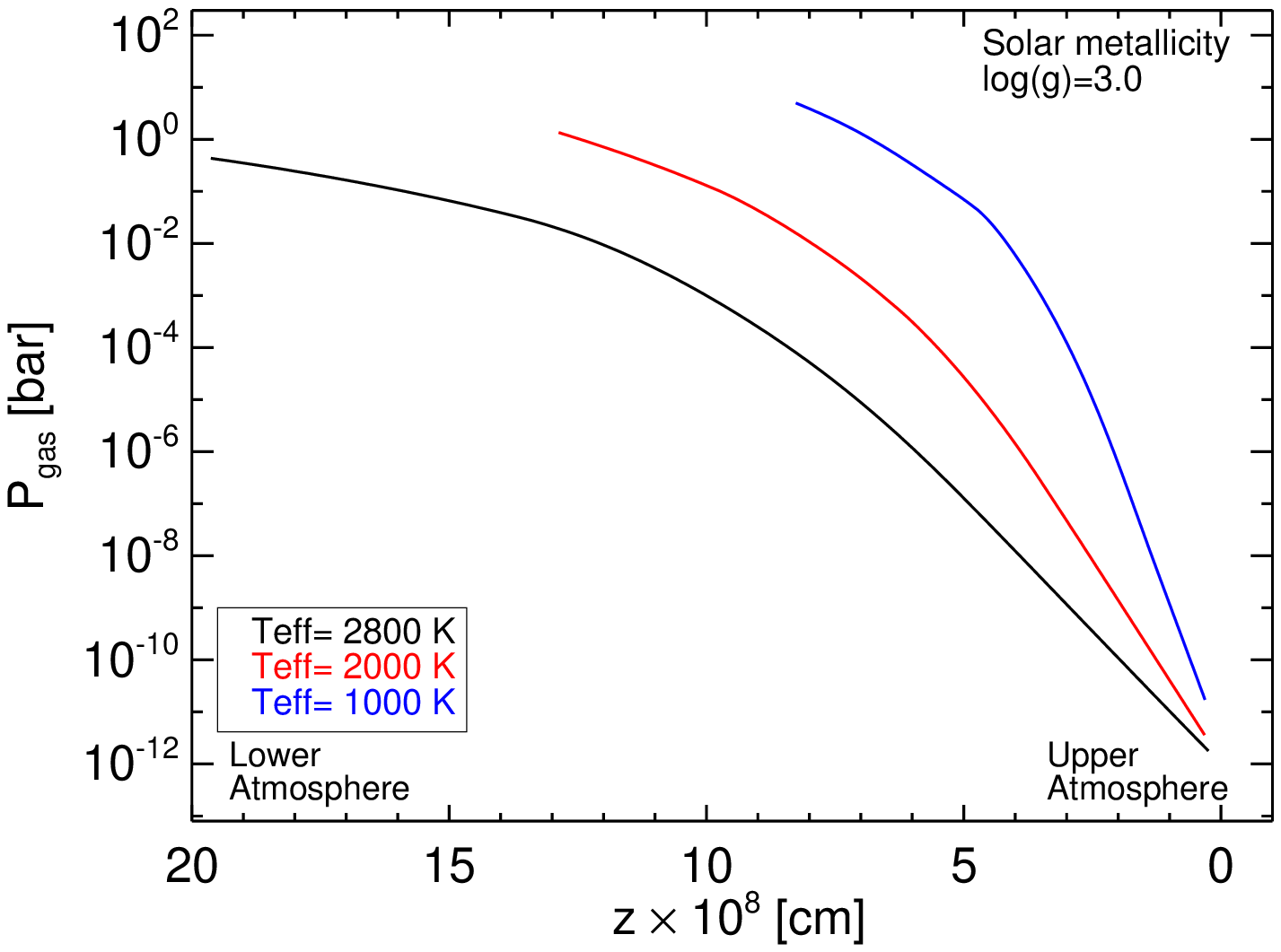}\\*[-0.8cm]
\hspace*{-1cm}\includegraphics[angle=0,width=0.6\textwidth]{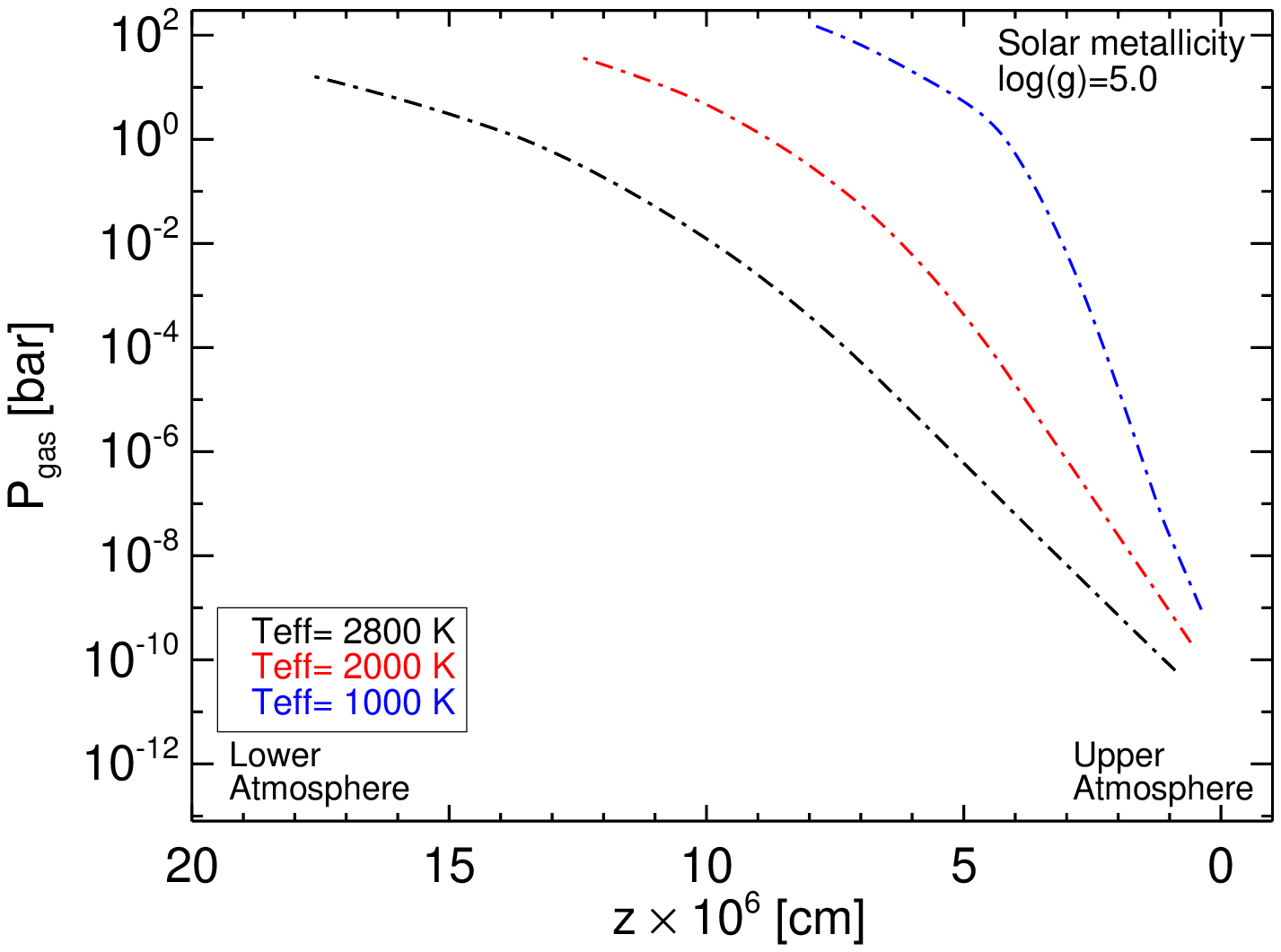}
\\*[-0.6cm]
\vspace{+0.1cm}\caption{Local gas pressure, P$_{\rm gas}$ as function of vertical geometrical extension of the atmosphere, z [cm] given by {\sc Drift-Phoenix}. {\bf Top:} Young brown dwarf atmospheres {\bf Bottom:} Old brown dwarf atmospheres.}\label{fig:Pgasz}
\end{figure}

\end{document}